\documentclass[11pt]{article}

\usepackage{graphicx}
\usepackage{amssymb}
\usepackage[figuresright]{rotating}
\usepackage{subfigure}
\usepackage{amsmath}
\usepackage{hyperref}
\usepackage{multirow}
\usepackage{enumerate}
\usepackage{color}
\usepackage{tabu}
\usepackage{natbib}    % 	for formatting citations in text
\usepackage{JASA_manu}
\usepackage{threeparttable}
\usepackage{chngcntr}
\usepackage{setspace}
\usepackage{mdwlist}
\usepackage{diagbox}
\usepackage{rotating}
\usepackage{float}
\usepackage{setspace}
\usepackage{comment}
\newcommand{\tabincell}[2]{\begin{tabular}{@{}#1@{}}#2\end{tabular}}

\begin{document}

\title{Simultaneous Estimation of Number of Clusters and Feature Sparsity in Clustering High-Dimensional Data}
\author{Yujia Li\\
Department of Biostatistics \\ University of Pittsburgh, Pittsburgh, PA 15261  \\ 
email: \texttt{yul178@pitt.edu} 
\and
Xiangrui Zeng  \\
Department of Computational Biology \\ Carnegie Mellon University, Pittsburgh, PA 15213  \\ 
email: \texttt{xiangruz@andrew.cmu.edu} 
\and
Chien-Wei Lin  \\
Division of Biostatistics \\ Medical College of Wisconsin, Wauwatosa, WI 53226  \\ 
email: \texttt{chlin@mcw.edu} 
\and
George Tseng \\
Department of Biostatistics \\ 
University of Pittsburgh, Pittsburgh, PA 15261\\
email: \texttt{ctseng@pitt.edu} 
}

\maketitle

\newpage

\mbox{}
\vspace*{2in}

\begin{center}

\textbf{Author's Footnote:}
\end{center}
Yujia Li is PhD student at Department of Biostatistics, University of Pittsburgh, Pittsburgh, PA 15261(email: yul178@pitt.edu).
Xiangrui Zeng is PhD student in Computational Biology, Carnegie Mellon University, PA 15213(email: xiangruz@andrew.cmu.edu). 
Chien-Wei Lin is an Assistant Professor in the Division of Biostatistics at the Medical College of Wisconsin, WI 53226(email: chlin@mcw.edu). 
George Tseng is Professor at Department of Biostatistics (primary appointment), Department of Human Genetics, Department of Computational Biology, University of Pittsburgh, Pittsburgh, PA, 15261(email: ctseng@pitt.edu).

\newpage
\begin{center}
\textbf{Abstract}
\end{center}
Estimating the number of clusters ($K$) is a critical and often difficult task in cluster analysis. Many methods have been proposed to estimate $K$, including some top performers using resampling approach. When performing cluster analysis in high-dimensional data, simultaneous clustering and feature selection is needed for improved interpretation and performance. To our knowledge, none has investigated simultaneous estimation of $K$ and feature selection in an exploratory cluster analysis. In this paper, we propose a resampling method to meet this gap and evaluate its performance under the sparse $K$-means clustering framework. The proposed target function balances between sensitivity and specificity of clustering evaluation of pairwise subjects from clustering of full and subsampled data. Through extensive simulations, the method performs among the best over classical methods in estimating $K$ in low-dimensional data. For high-dimensional simulation data, it also shows superior performance to simultaneously estimate $K$ and feature sparsity parameter. Finally, we evaluated the methods in four microarray, two RNA-seq, one SNP and two non-omics datasets. The proposed method achieves better clustering accuracy with fewer selected predictive genes in almost all real applications.

\vspace*{.3in}

\noindent\textsc{Keywords}: {Feature selection, 
$K$-means, 
Number of clusters, 
Sparse-$K$means}

\newpage

\section{Introduction}
\label{sec:Introduction}

Cluster analysis, an essential tool for unsupervised machine learning, is a set of useful data mining techniques to identify groups of objects of similar pattern. After dissimilarity structure is defined for every pair of objects (e.g. by Euclidean distance, Gower’s distance or 1 minus Pearson correlation), many clustering algorithms such as $K$-means \citep{macqueen1967some}, $K$-medoids \citep{kaufman1987clustering}, hierarchical clustering \citep{johnson1967hierarchical} and model-based clustering \citep{banfield1993model}, can be applied. In $K$-means and many other globally optimized algorithms, determination of the number of clusters $K$ is a critical, yet difficult task that needs to be pre-estimated before implementation. In the literature, many methods have been developed for this purpose in the $K$-means framework (e.g. the NbClust R package\citep{charrad2014package} provides implementation of 30 indices for this purpose) and they generally fall into two categories: estimation by cluster tightness or by resampling evaluation. In Section \ref{sec:EstimateKsummary} and \ref{sec:EstimateKresample}, we will outline different types of approaches in these two categories and justify the choice of methods we will evaluate in this paper. As expected, cluster analysis is an exploratory tool and often not mathematically well-defined under complex data structure. Thus, the best method to determine the number of clusters can vary depending on the data characteristics and purpose of clustering. Our purpose is to seek one or several top performers acrosswide variety of settings.

In modern data science, high-dimensional data with only moderate sample size are becoming more and more prevalent, which leads to small-n-large-p problems. Taking microarray or RNA-seq data as an example, $p$ is usually at the scale of 2,000 $\sim$ 20,000 and $n$ is only at 50$\sim$500. In cluster analysis with very large $p$, it is generally believed that only a small fraction of features are informative whereas the other features are essentially random noise which may disrupt discovery of the true cluster structure. Methods for clustering with feature selection has drawn increasing attention in the field \citep{witten2010framework,pan2007penalized,zhou2009penalized}. Take sparse $K$-means method \citep{witten2010framework} as an example, it transforms the $K$-means target function into maximizing the between-cluster sum-of-squares (BCSS) and imposes an $L_1$ penalty on gene weights into a weighted $K$-means framework to facilitate feature selection (see Section 4.1 for more details). In general, adequate feature selection in high-dimensional clustering not only improves clustering accuracy but also enhances model interpretation. To implement methods for clustering with feature selection in real applications, it requires simultaneous estimation of $K$ and $\lambda$, where $\lambda$ is the feature selection penalization parameter. Under Gaussian mixture model setting, \cite{pan2007penalized} applied a BIC criterion to select $K$ and $\lambda$. The strong Gaussian assumption, however, makes it vulnerable to non-Gaussian distributions and outliers in real applications.  \cite{witten2010framework} assumed $K$ is known and applied gap statistic to estimate $\lambda$ for the sparse $K$-means method.  To the best of our knowledge, an in-depth evaluation of simultaneous estimation of $K$ and $\lambda$ has not yet been investigated in the field. 
%{\color{red}{It mentioned in real application section that gap statistic will seemingly overestimate the number of features with non-zero weights when evaluating SNP data, relecting the need for a more accurate method to estimate the parameters}}.

In this paper, we take $K$-means and sparse $K$-means as clustering engine and propose a resampling based framework, which can estimate $K$ for $K$-means and estimate $K$ and $\lambda$ simultaneously for sparse $K$-means. Our framework is a stability-based subsampling method, which balances both sensitivity and specificity in the evaluation and is named as S4 (\underline{S}ubsampling \underline{S}core considering \underline{S}ensitivity \& \underline{S}pecificity) method. The article is structured as follows. In Section \ref{sec:Existingmethod}, we review $K$-means method and existing methods to estimate $K$ in the cluster analysis without feature selection. Our proposed S4 method for estimating $K$ is then introduced in Section \ref{sec:Proposes4}. In Section 4, we review clustering with feature selection using sparse $K$-means and introduce S4 method for simultaneous estimation of $K$ and $\lambda$. Section \ref{sec:simulations} includes extensive simulation results for clustering without and with feature selection under wide variety of data characteristics. Section \ref{sec:Application} contains results of large-scale real applications of 9 high-dimensional datasets. Finally, Section \ref{sec:Discussion} provides final conclusion and discussion. Our contributions are three-fold: 1) we extend existing methods (such as gap statistics and prediction strength) for estimating $K$ to simultaneously estimate $K$ and $\lambda$. 2) we introduce a simple, yet effective S4 method to simultaneously estimate K an $\lambda$. 3) we conduct extensive simulations and real applications to identify the best performers. In the all evaluations, S4 is always among the top performers.
%\label{eq:$K$-means-WCSS} 

\section{Existing methods for estimating $K$ in $K$-means (without feature selection)}
\label{sec:Existingmethod}
$K$-means algorithm \citep{macqueen1967some} is a popular clustering method due to its simplicity and fast computation. Suppose the data matrix is denoted as $X_{n\times p}$, where n is the number of subjects and p is total number of features and we want to cluster the data into $K$ clusters. $K$-means algorithm aims to minimize the within cluster sum of square (WCSS): 
\begin{equation}
\label{eqn:Kmeansequation}
\min \limits_{C} \sum\limits_{j=1}^p WCSS_j (C)=\min\limits_{C}\sum\limits_{j=1}^p \sum\limits_{k=1}^K \frac{1}{n_k}\sum\limits_{i_1,i_2 \in C_k} d_{i_1i_2,j}
\end{equation}
where $C=(C_1, C_2, \cdots, C_K) $ denotes the clustering results in which the $K$-means algorithm assigns all the samples into K partitions, $n_k$ denotes the number of samples in the $k$-th partition and $d_{i_1i_2,j}$ denotes the squared Euclidean distance of feature $j$ between sample $i_1$ and sample $i_2$. 

The number of clusters $K$ has to be estimated \textit{a priori}.  Many methods have been proposed for this purpose, and they generally can be divided into two categories: estimation by cluster tightness or by resampling evaluation. Table \ref{Methodintro} outlines the two categories of methods and more details are provided in the next subsection.

		\begin{table}
			%\begin{table}
			
			\begin{center}
				\caption{Summary of existing methods for estimating $K$ in $K$-means}\label{Methodintro}
				
			\end{center}
			%\tabincell{c}{extension for simultaneous  estimation of $K$ and $\lambda$}
			\resizebox{\textwidth}{!}{
				\begin{tabu}{|c|c|c|c|c|}
					\hline
					\multirow{4}*{Category}&\multirow{4}*{Subcategory}&\multirow{4}*{Abbreviation}&\multirow{4}*{Reference} &\multirow{4}*{\tabincell{c}{Extension for \\simultaneous  \\estimation\\ of $K$ and $\lambda$}}\\
					&                                          &                                           &                                       &\\
					&                                          &                                           &                                        &\\
					&                                           &                                           &                                       &\\
					% Category&Subcategory&Abbreviation&Reference&Extensible\\
					\cline{1-5}
					\multirow{6}*{\tabincell{c}{Cluster \\tightness}}                         &\multirow{4}*{summary index} &Silhouette*  & \cite{rousseeuw1987silhouettes}& \\
					&                                                 &CH*&\cite{calinski1974dendrite}&\\
					&                                                    &H index&\cite{hartigan1975clustering}&\\
					&                                               &KL*&\cite{krzanowski1988criterion}&\\
					\cline{2-5}
					&\multirow{2}*{Gap Statistic}   & GapPCA*&\multirow{2}*{\cite{tibshirani2001estimating}}&\multirow{2}*{\checkmark}   \\
					&                                               &GapUnif*&                          &\\
					\cline{2-5}                                                                      
					&Jump statistic                          & Jump*  & \cite{sugar2003finding}&  \\
					\cline{1-5}
					\multirow{7}*{Resampling}                 	 &  \multirow{4}*{stability-based}     & LD*&\cite{levine2001resampling}&\\     
					&                                                    &Ben-Hur    &\cite{ben2001stability}&\\
					&                                                     &FW*              &\cite{fang2012selection}&\\
					                                               &                                                     &S4*                    &                                   &\checkmark\\
					                                               \cline{2-5} 
					& \multirow{3}*{prediction-based}&PS*&\cite{tibshirani2005cluster}&\checkmark\\ 
					&                                                  &Clest&\cite{dudoit2002prediction}&\\      
					&                                                   &Lange &\cite{lange2004stability}&\\

					%	\hline
					%	&RSS & 3 & 0.96 & 90   \\
					%	Resampling&gap statistic & 3 & 0.83 &4906      \\
					%	&Prediction Strength& 4  & 0.83 & 70 \\
					\cline{1-5}
					
			\end{tabu}}
			\begin{center}
				\begin{tablenotes}
					
					\item*Methods included for extensive comparision in this paper.
					%\item[2] The last column indicates whether the methods are appropriate to be extended to estimate sparse parameter for sparse kmeans
				\end{tablenotes}
			\end{center}
			%\end{table}
			%\label{table:existingmethod} 
		\end{table}
\subsection{Estimation by Cluster Tightness}
\label{sec:EstimateKsummary}
As shown in Table \ref{Methodintro}, many classical methods for determining $K$ are based on cluster tightness using within cluster dispersion $W_K=\sum\limits_{j=1}^p WCSS_j (C_K)$, where $C_K$ is the output clustering result given $K$. The within cluster dispersion $W_K$ is a decreasing function with respect to $K$ and the underlying true $K$ is usually reflected as an elbow point. Specifically, $W_K$ initially drops quickly and the decrease flattens markedly after the underlying true $K$ (See Supplement Figure S1). Detection of such an elbow point in real data is often subjective and difficult. Many estimation methods depend on an index or transformation of $W_K$ to amplify the signal and capture the elbow point by optimization or certain decision rules. For example, \cite{calinski1974dendrite} proposed CH index to select $K$ to maximize $\frac{BCSS(k)/(k-1)}{WCSS(k)/(n-k)}$, where n is sample size. \cite{milligan1985examination} performed a comprehensive comparison of 30 variety of indexes and concluded that the CH index was one of the best performers. \cite{krzanowski1988criterion} proposed a KL index by maximizing $|\frac{DIFF(k)}{DIFF(k+1)}|$, where $DIFF(k)=(k-1)^{2/p}W_{k-1}-k^{2/p}W_k$ and p is the number of features. \cite{hartigan1975clustering} proposed H index by calculating $H(k)=\{\frac{W(k)}{W(k+1)}-1\}\times(n-k-1)$ and then $K$ is estimated as the smallest k such that $H(k)\leq 10$. \cite{rousseeuw1987silhouettes} developed silhouette index by maximizing $\frac{b(i)-a(i)}{max\{a(i),b(i)\}}$, where a($i$) is the average dissimilarity between subject $i$ and all other subjects in the cluster to which subject $i$ belongs and b($i$) is the smallest average dissimilarity of $i$ to all points in any other cluster, of which $i$ is not a member. As shown in Table \ref{Methodintro}, we include four well-known methods: CH index, KL index, H index and silhouette as representative summary indexes into our comparisons.

In addition to methods based on summary indexes, \cite{tibshirani2001estimating} proposed to maximize a gap statistic defined as the difference between the original $W_K$ and the null (reference) $W_K$ obtained from permutation where data do not contain cluster structure. Specifically, the gap statistic is $gap(K)= (\sum_{b=1}^B \log(W_K^{(b)}))/B-\log(W_K)$, where $W_K^{(b)}$ is the simulated null $W_K$ from uniform distribution or PCA rotation in the $b$-th simulation and $B$ is total number of simulations. Conceptually, subtracting the null $W_K$ from the observed $W_K$ serves to de-trend (or normalize) the decreasing pattern so that the true $K$ can be obtained by maximizing the gap statistic. \cite{sugar2003finding}, based on information theoretic perspective, later proposed a jump statistics by $jump(K)=(W_K)^y-(W_{K-1})^y$ where the transformation power $y$ is typically chosen as $-p/2$ and $p$ is the total number of features. We include gap and jump statistic in the comparison. 
\subsection{Estimation by Resampling Evaluation}
\label{sec:EstimateKresample}
Another category of methods to estimate $K$ is by resampling evaluation, including subsampling or bootstrap. With data perturbations introduced from resampling, clustering from different resampled data should generate stable (similar) results when the underlying true $K$ is selected.  \cite{levine2001resampling} proposed to measure the concordance between subsampled data and the original whole data to assess stability. \cite{ben2001stability} measured the stability across subsampled data and used the transition of distribution of similarity score to determine the optimal $K$. \cite{fang2012selection} compared pairwise bootstrapped data to examine the stability. We note that the Ben-Hur method is not completely quantitative since users need to manually check the transition of the distribution. Therefore, we choose the LD and FW methods as representatives of stability-based methods for comparisons. Our proposed S4 method also belongs to this category and is introduced in Section 3.

In contrast to stability-based methods, a group of methods split the original data into two portions, pretend the first portion as training data and the second portion as testing data and mimic supervised machine learning setting to evaluate prediction accuracy. The underlying true $K$ should generate the highest prediction accuracy. Take  \cite{tibshirani2005cluster} as an example, the method randomly splits data $X$ into training data $X_{tr}$ and testing data $X_{te}$. Training data are clustered into $K$ clusters (denoted as $C(X_{tr}, K)$), and the resulting $K$ cluster centroids are used as a classifier to assign test samples into $K$ clusters. The element ($i_1$, $i_2$) of co-membership matrix $D[C(X_{tr},k),X_{te}]_{i_1,i_2}=1$ if sample $i_1$ and $i_2$ of testing data are predicted in the same cluster by the training data centroids and 0 otherwise. By comparing  clustering results between testing data on training centroids and testing data on test clusters ($A_{k1},A_{k2},\cdots,A_{kk}$) for a given number of cluster $k$, the prediction strength for given k is defined as   
\begin{equation}
\label{eqn:Pslowdimension}
ps(k)=\min\limits_{1\leq j \leq k}\frac{1}{n_{kj}(n_{kj}-1)}\sum\limits_{i_1,i_2 \in A_{kj}} D[C(X_{tr},k),X_{te}]_{i_1,i_2}
\end{equation}
where $n_{k1},n_{k2},\cdots n_{kk}$ are the number of samples in clusters $A_{k1},A_{k2},\cdots,A_{kk}$. \cite{dudoit2002prediction} proposed clest method which uses reference data to adjust the prediction score. However, this method has been criticized to contain many unspecified parameters and hard to implement in practice \citep{lange2004stability}. \cite{lange2004stability} proposed a different framework to adjust prediction score by reference data. However, the method requires heavy computation to measure prediction score for both original and repeatedly simulated reference data and no software package was provided for implementation. Thus, only the prediction strength (PS) method from this category is selected for comparison in this paper.

\section{Proposed S4 method without feature selection}
\label{sec:Proposes4}
We here propose a stability-based resampling method called S4 (\underline{S}ubsampling \underline{S}core considering \underline{S}ensitivity \& \underline{S}pecificity) by measuring the stability of clustering in repeated subsampled data. The clustering result of $X_{n\times p}$ can be presented by an $n\times n$ comembership matrix $T$ where $T_{i,j}$ indicates whether subject $i$ and subject $j$ are clustered together in the same cluster.
%as shown in equation \ref{eqn:Clustercomembership}.
\begin{eqnarray}
\label{eqn:Clustercomembership}
T_{ij}=\left\{
\begin{array}{rcl}
\begin{split}
1      &      & \text{subjects $i$ and $j$ belong to the same cluster}\\
0     &      & \text{otherwise} \\
\end{split}
\end{array} \right. 
\end{eqnarray}
We generate $B$ sets of subsampled data, denoted as $X^{(1)}_{n\times p}, X^{(2)}_{n\times p},\cdots,X^{(B)}_{n\times p}$, by randomly selecting $f$ ($0<f<1$) fraction of the original data and the $ n \times n$ comembership matrix for the $b$-th subsample is denoted as $T^{(b)} $($b=1, 2, \cdots, B$) defined by (3), where value NA is assigned if one or both of the two subjects $i$ and $j$ are not selected in the $b$-th subsample. We take element-wise average of $T^{(1)},T^{(2)},\cdots,T^{(B)}$ to derive the mean comembership matrix $\bar{T}^{(sub)}$ from all subsampled data, where $\bar{T}^{(sub)}_{i,j}$ indicates the proportion of times subject $i$ and subject $j$ are clustered together across all $B$ subsamples. Missing data are omitted during the averaging process.

The S4 method measures concordance between comembership matrix of the original data $T$ and the averaged comembership matrix from repeated subsampled data $\bar{T}^{(sub)}$. Given the number of clusters $K$, the concordance score for the $i$-th subject, is defined as follows:
\begin{eqnarray}
\label{eqn:Clustercomembership}
S_i(K)=
\begin{array}{rcl}
\begin{split}
\frac{\sum\limits_{j\neq i} \bar{T}^{(sub)}_{i,j}\cdot I\{T_{i,j}=1 \}}{\sum\limits_{j\neq i} I\{T_{i,j}=1 \}}+\frac{\sum\limits_{j\neq i} (1-\bar{T}^{(sub)}_{i,j})\cdot I\{T_{i,j}=0 \}}{\sum\limits_{j\neq i} I\{T_{i,j}=0 \}}-1,     &      & i=1, 2, \cdots, n\\
\end{split}
\end{array} 
\end{eqnarray}
where $I(\cdot)$ is an indicator function that takes value 1 if the statement is true and 0 otherwise. If we treat the comembership matrix $T$ as the underlying truth and borrow the notion of supervised machine learning, the first term $\frac{\sum\limits_{j \neq i} \bar{T}^{(sub)}_{i,j}\cdot I\{T_{i,j}=1 \}}{\sum\limits_{j\neq i} I\{T_{i,j}=1 \}}$ can be regarded as sensitivity score of sample $i$ and the second term $\frac{\sum\limits_{j \neq i}(1-\bar{T}^{(sub)}_{i,j})I\{T_{i,j}=0 \}}{\sum\limits_{j \neq i}\cdot I\{T_{i,j}=0 \}}$ can be considered as specificity score of sample $i$. The $S_i$ score is equivalent to Youden index (sensitivity+specificity- 1) \cite{youden1950index} for evaluating a dichotomous diagnostic test.

Scattered points (i.e. subjects which are randomly scattered and not clearly close to any cluster center) may severely interfere the performance of resampling-based methods \citep{tseng2007penalized,maitra2009clustering}. In our definition of $S_i$, the value should be close to 1 when subject $i$ is a stably clustered subject but approach 0 if subject $i$ is a scattered point. To avoid the impact of potentially scattered subjects, we define $S^*_{\rho}(K)$ as the trimmed mean of $S_i(K), 1\leq i\leq n$ by truncating the lower $\rho\%$ of subjects. Conceptually, we estimate $K$ that maximizes $S^*_{\rho}(K)$ with $\rho$ pre-defined. In practice, we develop an iterative approach to drop $\rho\%$ subjects. Suppose the concordance score for each sample in the beginning is $S^0_{1}(K)$, $S^0_{2}(K)$, $\cdots$, $S^0_{n}(K)$ for n subjects and K clusters, we first find the most unstable subject with lowest $S^0_{i}(K)$ and drop it, then we recalculate the concordance score for all the remaining samples to get $S^1_{1}(K)$, $S^1_{2}(K)$, $\cdots$, $S^1_{n-1}(K)$ by equation (4) using $\bar{T}^{(sub)}_{n-1,n-1}$ and $T_{n-1,n-1}$ after dropping the subject in $\bar{T}^{(sub)}_{n,n}$ and $T_{n,n}$. We drop the second subject with lowest concordance score from $S^1_{1}(K)$, $S^1_{2}(K)$, $\cdots$, $S^1_{n-1}(K)$. This iterative way continues until $\rho \%$ subjects are dropped and then the average concordance score for all the samples after dropping $\rho\%$ is calculated as $S^*_{\rho}(K)$. 

Figure S2 shows a toy example of three clusters in two dimensions where two clusters are closer to each other compared to the third one. In this case, $\hat{K}=2$ and $\hat{K}=3$ both show perfect score: $S^*_0(2)=S^*_0(3)=1 $, where $\rho=0$. When multiple $K$ generate the same highest $S^*$ score, we will take the larger $K$ as the solution (i.e. $\hat{K}=3$ in this case). Figure S3-S5 shows the ordered concordance score for Simulation setting 2-4 when K varies from 2 to 5 (the detail of simulation settings will be shown in Section 5.1). It is clearly that a few of points (scattered points) have much lower score compared with the majority. These points needs to be trimmed since what we care more is the remaining majority . In Section 5.2, Table 3 shows the benefit of trimming in these three settings.

In the case of detecting null data (i.e. $K=1$), it remains a challenging problem for resampling methods since all subjects are in one cluster. In other words, $K=1$ always achieves the highest stability. \cite{tibshirani2005cluster} proposed to select $K=1$ if $ps(K)+se(K)<0.8$ for all $2\leq K\leq K_{max}$, where $se(K)$ is the standard error of the prediction strength at $K$ over cross validation. We borrow this idea to choose $K=1$ when $S^*_{\rho}(K)$ is less than a threshold $s_0$ for all $2\leq K\leq K_{max}$. 

The step-by-step algorithm of S4 method is as follows:

\leftmargini=8mm
\begin{itemize*}
	\item[Step1]: Given an original data $X_{n \times p}$ with n samples and p features, for every $K$ ($2\leq K\leq K_{max}$), generate $B$ sets of randomly subsampled data with sample size $f\times n$ ($0<f<1$), taken as interger. Denote subsampled datasets as $X^{(1)}, X^{(2)},\cdots,X^{(B)} $. 
	\item[Step2]: Perform $K$-means to the original data and obtain comembership matrix $T$.
	\item[Step3]: Perform $K$-means to each subsample dataset to calculate comembership matrix $T^{(1)},T^{(2)},\cdots,T^{(B)}$. Derive $\bar{T}^{(sub)}$ by taking average across $B$ comembership matrices.
	\item[Step4]: Calculate the concordance score for every sample, $S_1(K),S_2(K),\cdots,S_n(K)$ based on Equation (\ref{eqn:Clustercomembership}). Derive the trimmed mean $S^*_{\rho}(K)$ by removing the lower $\rho\%$ of samples. Define $S_{max}=\max_{2\leq K\leq K_{max}} S^*_{\rho}(K)$. 
	\item[Step5]: If $S_{max}<s_0$, return $\hat{K}=1$. If $S_{max}\geq s_0$, return $\hat{K}=\arg\max_{2\leq K\leq K_{max}} S^*_{\rho}(K)$. 
\end{itemize*}
Throughout this paper, we set $\rho=5$ and $s_0=0.8$. The selection of parameters is justified by extensive simulations in Section 
\ref{sec:simulations}.

\section{Simultaneous estimation of $K$ and $\lambda$ when clustering with feature selection}
\subsection{Review of Sparse $K$-means} 
\label{sec:Sparsekmeans}
In $K$-means algorithm, every feature is used and equally weighted in the distance derivation. In high-dimensional data, many irrelevant features exist and may interfere with detection of true cluster structure. For example, when we cluster samples in gene expression data to detect potential disease subtypes, incorporating feature selection in the clustering analysis not only improves clustering accuracy but also provide biological interpretation as to which features (genes) contribute to characterize the disease subtypes. \cite{witten2010framework} proposed a sparse $K$-means approach with lasso regularization on feature-specific weights to tackle this problem. Consider to extend minimizing within cluster sum of square (WCSS) in $K$-means to weighted WCSS with $L_1$ regularization of features. The new objective function: $\min\limits_{C,w} \sum\limits_{j=1}^pw_j \times WCSS_j$ with regularization on weights unfortunately leads to a null solution with all weights diminishing to zero. Thus, \cite{witten2010framework} converted the problem of minimizing weighted WCSS into maximizing weighted Between Cluster Sum of Squares(BCSS) using the fact that total sum of squares $TSS=BCSS+WCSS$ and $TSS$ is a constant. The final objective function in sparse $K$-means becomes:
\begin{eqnarray}
\label{eqn:Sparsekmeans}
\begin{split}
& \max\limits_{C,w}\sum\limits_{j=1}^p w_j \times BCSS_j (C)=\max\limits_{C,w} \sum\limits_{j=1}^pw_j \times\bigg[\frac{1}{n}\sum\limits_{i_1,i_2}d_{i_1i_2,j}-\sum\limits_{k=1}^K \frac{1}{n_k}\sum\limits_{i_1,i_2 \in C_k} d_{i_1i_2,j}\bigg]\\
& subject \quad to \quad ||w||_2\leq 1,||w||_1 \leq \lambda, w_j \geq 0 \quad \forall j\\
\end{split}
\end{eqnarray}
where $w=(w_1, \cdots, w_p)$, $w_j$ is the weight for feature $j$, $C=(C_1,C_2,\cdots,C_K)$, and $||w||_1$ and $||w||_2$ are $L_1$ and $L_2$ norms of weight $w$. The $L_1$ regularization shrinks most of the features to 0 weight while performing clustering. The number of cluster $K$ and sparsity parameter $\lambda$ must be estimated \textit{a priori}. \cite{witten2010framework} assumed that $K$ is pre-estimated and they used gap statistic to estimate $\lambda$.

To the best of our knowledge, methods for simultaneously estimating $K$ and $\lambda$ have not been systematically developed in the literature. To fill this gap, we extend gap statistic, prediction strength and S4 method in the next subsection for simultaneously estimation of $K$ and $\lambda$.

\subsection{Extension of S4, Gap Statistic and Prediction Strength}
\label{sec:Extension}
%one approach is to estimate $K$ and $\lambda$ by optimizing the sum of the two scores to achieve performance of both:
%\vspace{-0.1cm}
%\begin{equation*}
%(K^*,\lambda^*)= \arg \max\limits_{K, \lambda} S(K,\lambda)+T(K,\lambda)
%\end{equation*}

\textit{\underline{S4:}} To apply S4 method to simultaneously estimate $K$ and $\lambda$ in sparse K-means, for every pair of parameters $(K,\lambda)$, we first caluclate cluster concordance score of every subject $S_i(K,\lambda)$ by equation (3) in Section \ref{sec:Proposes4}. Similarly, we take the trimmed mean statistic $S^*_{\rho}(K,\lambda)$ of $S_i(K,\lambda)$, $1 \leq i \leq n$, by removing $\rho$ \% of samples with lowest concordance score iteratively. Second, we  also define a feature selection concordance score $F(K,\lambda)$. Denote by $f_j$ the feature selection index for feature $j$ in clustering of the original data (i.e. $f_j=1$ if feature $j$ is selected and $f_j=0$ otherwise). Similarly, $f_j^{(b)}$ is the feature selection index for feature $j$ in the $b$-th subsampling and $f_j^{sub}= (\sum_{b=1}^B f_j^{(b)})/B$ the proportion that feature $j$ is selected among $B$ subsamplings. We define the feature selection concordance score as
\vspace{-0.1cm}
\begin{equation*}
F(K, \lambda)=\frac{\sum _{j=1}^p f^{sub}_{j}I\{f_{j}=1 \}}{\sum_{j=1}^p I\{f_{j}=1 \}}+\frac{\sum_{j=1}^p(1- f^{sub}_{j})I\{f_{j}=0 \}}{\sum_{j=1}^p I\{f_{j}=0 \}}-1,
\end{equation*}

\noindent where the first term represents feature selection sensitivity when clustering result of the original dataset is treated as the underlying truth, the second term is specificity and $F(K,\lambda)$ takes the form of Youden index in supervised machine learning. 

After the clustering $S^*_{\rho}(K, \lambda)$ and feature selection $F(K, \lambda)$ concordance scores are defined, one naïve approach is to estimate $K$ and $\lambda$ by optimizing sum of the two concordance scores:
\vspace{-0.1cm}
\begin{equation*}
(\hat{K},\hat{\lambda})= \arg \max\limits_{K, \lambda} S^*_{\rho}(K,\lambda)+F(K,\lambda)
\end{equation*}

This solution, however, does not generate desirable result even in simple simulations (see Supplement Simulation S1). This is mainly because the roles and behavior of $K$ and $\lambda$ are quite different. $K$ is discrete, is more stable in the optimization and has more critical impact to the final clustering. On the other hand, $\lambda$ is continuous and slight change of $\lambda$ usually does not significantly alter the clustering result. Hence, we propose an alternative two-stage approach by first focusing on clustering concordance score to select $K$ since the instability of the feature concordance scores can greatly impact the selection of $K$ (see Supplement Simulation S1). $\lambda$ is then selected by the sum of clustering and feature concordance scores. Specifically, we first obtain $\hat{K}$ by
\vspace{-0.1cm}
\begin{equation*}
\hat{K}= \arg\limits_{K} \max\limits_{K,\lambda} S^*_{\rho}(K,\lambda).
\end{equation*}

Next, given $\hat{K}$, we then estimate $\hat{\lambda}$ by
\vspace{-0.1cm}
\begin{equation*}
\hat{\lambda}= \arg \max\limits_{\lambda} S^*_{\rho}(\hat{K},\lambda)+F(\hat{K},\lambda)
\end{equation*}

Remark: In the definition of feature selection concordance score, we always drop the case when all features are selected in the optimization since the denominator of the second term (specificity) is zero and the score is therefore not well-defined. This restriction generally has limited impact in the final result.

\noindent \textit{\underline{Gap statistic: }} We next extend gap statistic for the joint estimation of $K$ and $\lambda$. For given ($K$, $\lambda$), suppose $w^*=(w_1^*,\cdots,w_p^*)$ and $C^*=(C_1^*,\cdots,C_K^*)$ are the solution of sparse K-means from equation (\ref{eqn:Sparsekmeans}). Consider the resulting target function:

\begin{equation*}
O(K, \lambda)=\sum\limits_{j=1}^p w_j^*(\frac{1}{n}\sum\limits_{i_1=1}^n \sum\limits_{i_2=1}^n d_{i_1i_2,j}-\sum\limits_{k=1}^K\frac{1}{n_k}\sum\limits_{i_1,i_2\in C_k^*} d_{i_1i_2,j}).
\end{equation*}

\noindent The bivariate gap statistic for a given ($K$, $\lambda$) can be defined as
\begin{equation*} 
Gap(K, \lambda)=log(O(K, \lambda))-\frac{1}{B}\sum\limits_{b=1}^Blog(O^{(b)}(K, \lambda)),
\end{equation*}
where $O^{(b)}(K, \lambda)$ is the target function from permuted reference data. We first choose the optimized ($\hat{K}, \lambda^{'}$) $= \arg \max\limits_{K, \lambda} Gap(K,\lambda)$. Given $\hat{K}$, The estimated $\hat\lambda$ is chosen as the smallest $\lambda$ whose gap statistic is within one standard deviation of the largest gap statistic: 
\begin{equation*}
(\hat{K},\hat{ \lambda})=\arg\min_{\lambda} \{\lambda: Gap(\hat{K}, \lambda) \geq [Gap(\hat{K}, \lambda^{'})-SD(Gap(\hat{K}, \lambda^{'}))] \}.
\end{equation*}
where $SD(Gap(\hat{K}, \lambda^{'}))$ is the standard deviation for gap statistic at $K=\hat{K}, \lambda=\lambda^{'}$. The final estimation output is ($\hat{K}, \hat{\lambda}$).
%The estimation is obtained by $(K^*, \lambda^*)=\arg\max Gap(K, \lambda)$.

\noindent \textit{\underline{Prediction strength:}} Finally, we also extend prediction strength (\cite{tibshirani2005cluster}) to simultaneously estimate $K$ and $\lambda$ for sparse $K$means:
\begin{equation*}
ps(K,\lambda)=\min\limits_{1\leq j \leq k}\frac{1}{n_{k\lambda j}(n_{k\lambda j}-1)}\sum\limits_{i_1,i_2 \in A_{k\lambda j}} D[C(X_{tr},k),X_{te}]_{i_1,i_2}
\end{equation*}
All the notation are similar to those of prediction strength in Section \ref{sec:EstimateKresample}. The only difference lies in $D[C(X_{tr},k),X_{te}]_{i_1,i_2}$. When using training data centroid to predict test samples, instead of using Euclidean distance, here we use weighted Euclidean distance and the weights are obtained by the result of sparse $K$-means of the training data. In addition, we use the features selected by training data to compare predictive features of testing data and measure the concordance of features. Following the similar rationale of S4, denote by $f^{(tr)}_j$ as the feature selection index for feature $j$ of the training data (i.e. $f_j=1$ if feature $j$ is selected otherwise $f_j=0$). Similarly, define $f_j^{(te)}$ the feature selection index for feature $j$ from test data. We define the feature prediction strength as
\vspace{-0.1cm}
\begin{equation*}
F_{ps}(K, \lambda)=\frac{\sum _{j=1}^p f^{(tr)}_{j}I\{f_{j}^{(te)}=1 \}}{\sum_{j=1}^p I\{f_{j}^{(te)}=1 \}}
\end{equation*}
Similar to S4, we first estimate $K$ by $\hat{K}= \arg\limits_{K} \max\limits_{K,\lambda} ps(K,\lambda)$. Given $\hat{K}$, $\hat{\lambda}= \arg \max\limits_{\lambda} ps(\hat{K},\lambda)+F_{ps}(\hat{K},\lambda)$.

\subsection{Efficient Algorithm for Choosing Grids of $\lambda$}
To achieve simultaneous optimization of $K$ and $\lambda$ in methods proposed in Section \ref{sec:Extension}, one difficulty in practice is to design adequate grid values for $K$ and $\lambda$. Since $K$ is discrete, we usually search from 2 to a pre-specified upper bound $\tilde{K}$ (e.g. $\tilde{K}=10$). For $\lambda$ in sparse K-means with pre-fixed $K$, a sequence of selected $\vec{\lambda}=(\lambda_1,\cdots,\lambda_m)$ monotonically correspond to a sequence of number of selected features $v_1, \cdots, v_m$ (i.e. if $\lambda_1\geq\lambda_2$, $v_1\geq v_2$). If the searching space $(\lambda_1,\cdots,\lambda_m)$ is not properly designed, much computing time will be wasted on $\lambda$ values that generate identical number of selected features. Below we propose a bisection algorithm to select $\vec{\lambda}=(\lambda_1,\cdots,\lambda_m)$ so that the corresponding numbers of selected features are roughly equally spaced in log2 scale. It is also obvious that the grid selection for $\lambda$ needs to depend on $K$ (i.e. we need $\vec{\lambda}_k=(\lambda_{k,1},\cdots,\lambda_{k,m_k})$ for $2\leq k\leq \tilde{K}$). Below is the detailed procedure to select $\lambda$ grids given $K$:
\begin{itemize}
	\item[1.] Initialize $\lambda_{k,1}^{(0)}=\lambda_0$ and $\lambda_{k,2}^{(0)}=\sqrt{p}$ in the 0-th iteration grids, where $\lambda_0$ is a small $\lambda$ that selects only a few features and $\sqrt{p}$ is the upper bound that guarantees to include all features.  
	\item[2.] Insert a new $\lambda$ by geometric mean $\lambda^{new}=\sqrt{\lambda_{k,1}^{(0)}\cdot\lambda_{k,2}^{(0)}}$ and form the 1st iteration of grids  \{$ \lambda^{(1)}_{k,1},\lambda^{(1)}_{k,2},\lambda^{(1)}_{k,3}$\}=\{$\lambda^{(0)}_{k,1},\lambda^{new},\lambda^{(0)}_{k,2} $\}. 
\item[3.] Derive the corresponding number of selected features $\{v^{(1)}_{k,1},v^{(1)}_{k,2},v^{(1)}_{k,3}\}$ using sparse $K$-means. 
\item[4.] Calculate $d_i=log(v^{(1)}_{k,i+1}) - log(v^{(1)}_{k,i})$, the log-scale intervals of existing grids. Select the largest log-scale interval $i^*=\arg\max\limits_i d_i$ and insert a new $\lambda^{new}=\sqrt{\lambda^{(1)}_{k,i^*}\cdot\lambda^{(1)}_{k,i^*+1}}.$
\item[5.] Sort \{$\lambda_{k,1}^{(1)}, \lambda_{k,2}^{(1)},\lambda_{k,3}^{(1)}, \lambda^{new}$\} to generate the 2nd iteration of grids $\{ \lambda^{(2)}_{k,1},\lambda^{(2)}_{k,2},\lambda^{(2)}_{k,3}, \lambda^{(2)}_{k,4}\}$. 
\item[6.] Repeat step 3, 4 and 5 up to the $m$-th iteration, which generates $m+2$ grids for $\lambda$: $\{ \lambda_{k,1}^{(m)}=\lambda_0, \lambda_{k,2}^{(m)}\cdots \lambda_{k,m+2}^{(m)}=\sqrt{p}\}$.
\item[7.] Delete any $\lambda$ grid that selects all features.
\end{itemize}

Using this process,  we obtain the searching grids of $K$ and $\lambda$ for optimization: \\$\bigg [ (1,\lambda_{1,1}), (1,\lambda_{1,2}), \cdots, (1,\lambda_{1,m_1}) \bigg ]$,  $\bigg [(2,\lambda_{2,1}), (2,\lambda_{2,2}), \cdots, (2,\lambda_{2,m_2})\bigg ]$, \\$\cdots \cdots$, $\bigg [ (\tilde{K},\lambda_{\tilde{K},1}), (\tilde{K},\lambda_{\tilde{K},2}), \cdots, (\tilde{K},\lambda_{\tilde{K}, m_{\tilde{K}}})\bigg ]$.

\section{Simulations}
\label{sec:simulations}
\subsection{Simulation I: Simulations for  Determining $K$ in $K$-Means}
\label{sec:Simulationlow}
%\subsubsection{simulation settings}
\textit{\underline{Simulation settings: }} We perform 10 different simulation settings including one null data setting (i.e. $K=1$; setting1), four well-separated simulation settings (settings 2-5), five non-well-separated simulation settings (settings 6-10). Well-separated settings and null data setting are replication of simulation study presented by \cite{tibshirani2001estimating} . For non-well-separated settings, setting 6 and setting 10 are directly from \cite{tibshirani2005cluster} and the other three settings are modified based upon these two. Detailed specification for all settings are outlined below:

%The null data setting and well-separated settings are illustrated below \citep{tibshirani2005cluster}.
\noindent \textit{Null data (K=1)}
\begin{itemize}
	\item Setting 1: We generate 200 samples from ten-dimensional uniform distribution, where each dimension ranges from 0 to 1.
\end{itemize}
\textit{Well-separated}
\begin{itemize}
	\item Setting 2: Three clusters in two dimensions are generated by standard normal distribution centering (0, 0), (0, 5) and (5, 3) respectively, with 25, 25 and 50 samples in each cluster.
	\item Setting 3: There are totally four clusters in three dimensions and the center for each cluster is randomly obtained by N(0, 5$\cdot I$). We use standard normal to randomly generate 25 or 50 observations for each cluster. If points of any two clusters have distance smaller than 1, we will discard this simulation and simulate the data again.
	\item Setting 4: There are totally four clusters in ten dimensions and the center for each cluster is randomly obtained by N(0, 1.9$\cdot I$). We use standard normal distribution to randomly generate 25 or 50 observations for each cluster. If points of any two clusters have distance smaller than 1, we will discard this simulation and simulate the data again.
	\item Setting 5: We simulate two clusters in three dimensions with 100 observations in each cluster. For the first cluster, choose $x_1=x_2=x_3=t$ where t is chosen by equal spaced values from -0.5 to 0.5, then add Gaussian  noise with standard deviation 0.1 to each feature. The second cluster are generated in the same way except for adding value 10 to each feature at the end. This forms two elongated cluster on main diagonal in three-dimensional cube.
\end{itemize}
\textit{Non-well-separated}
\begin{itemize}
	\item Setting 6: There are four clusters in two dimensions and each cluster is generated from standard normal distribution centering at (0,0), (0, 2.5), (2.5, 0), (2.5, 2.5), with 25 observations respectively.
	\item Setting 7: There are four clusters in two dimensions and each cluster is generated from standard normal distribution centering at (0,0), (0, 3), (3, 0), (3, 3), with 25 observations respectively.
	\item Setting 8: There are four clusters in two dimensions and each cluster is generated from standard normal distribution centering at (0, 0), (0, 3.5), (3.5, 0), (3.5, 3.5), with 25 observations respectively.
	%\item Setting 9: The first cluster is the same as first cluster in setting 5. For second cluster, instead of adding 10 to every feature, we only add value 0.5 to every feature. So there are two close and elongated clusters.
	\item Setting 9: Similar setting as setting 5 except for the second cluster. Instead of adding value 10 to each feature, we only add value 1 to each feature, producing two close and elongated clusters
	\item Setting 10: Similar setting as setting 5 except for the second cluster. Instead of adding value 10 to each feature, we only add value 1 to the first feature. Compared with setting 9, two clusters are closer to each other and are more difficult to separate.
	
\end{itemize}

All the simulation settings are repeated for 100 times and the searching space of number of clusters is chosen from 1 to 10. We compare S4 with nine existing methods described in Section \ref{sec:Existingmethod}. We perform $B=100$ resampling evaluation for all resampling-based method although our experience shows that $B=20$ already generates stable results.

%\linespread{2}

\noindent \textit{\underline{Sensitivity analysis for trimmed mean in S4: }} As described in the toy examples in Figure \ref{fig:Toyexample}, adequate trimming in cluster concordance scores before averaging could improve estimation performance. In Table \ref{table:Trimmedmeans}, we perform a sensitivity analysis of different trimming parameter $\rho=0, 2, 5, 8, 10, 15, 20$ by applying S4 to 10 simulation settings. We find that $\rho=2\sim 10$ generally provide good performance and $\rho=5$ seems to give the best overall estimation. Throughout this paper, we set $\rho=5$ in all comparisons.\\
\begin{table}
	%\centering
	%\renewcommand\arraystretch{2}	
	
	\caption {Simulation result for trimming different proportions of samples in clustering without feature selection. Each cell indicates how many times it chooses the correct $K$ in different settings by trimming different proportion of samples}\label{table:Trimmedmeans}
	\label{comp} 
	\hspace{-2cm}
	\begin{center}
		\begin{spacing}{1.19}
			%\resizebox{\textwidth}{!}{
			\resizebox{\textwidth}{30mm}{
			\begin{tabu}{|c|c|c|c|c|c|c|c|c|c|}
				\hline
				\diagbox{Setting}{$\rho \%$} & 0\% & 2\%& 5\%  & 8\% & 10\%  & 15\%& 20\% \\
				\hline
				setting1 $K=1$              & 100 & 99 & 98  & 98 & 97& 94& 90\\
				\hline
				setting2  $K=3$             & 74  & 96 & 100 &  100 & 100 & 96 & 95\\	
				\hline	
				setting3 $K=4$     & 70    & 90 & 99  & 99 & 99 & 98&97 \\	
				\hline
				setting4  $K=4$                     & 49    & 67 & 78  & 82 & 87& 95  &96 \\
				\hline
				%\hline
				setting5 $K=2$ & 100   & 100   &  91  & 57 & 37  & 3 &0 \\
				\hline
				setting6 $K=4$ & 40 & 43  & 40 & 42& 44  &47 & 49\\
				\hline
				setting7 $K=4$ & 69  & 70   &  70 & 70 & 73  & 79& 81 \\
				\hline
				setting8 $K=4$          & 82  & 83 &  79& 84 & 83 & 90& 97 \\
				\hline
				%Setting9               & 91 & 87 & 92  & 94 & 97& 100& 100\\
				%setting9               & 32  & 35 & 48 &  52 & 51 & 40 & 15\\		
				setting9  $K=2$    & 65    & 75 & 87  & 87 & 86 & 55&16 \\
				\hline	
				setting10  $K=2$                     & 0    & 3 & 4  & 6 & 3& 0  &0 \\
				\hline	
			\end{tabu}}
			\end{spacing}
			%}
	\end{center}
\end{table}
%\subsubsection{other approaches}
%The result for all 10 settings is listed in table 3. Every column indicates how many times each method chooses the correct $K$.
\noindent \textit{\underline{Comparison of ten methods in ten simulation settings: }} Table \ref{table:Lowdimensionsimulation} summarizes the number of correct $K$ estimation among 100 simulations using 10 methods in 10 simulation settings. We observe that gap statistic performs almost perfectly in well-separated settings 1$\sim$5, similar to what was shown in the original paper. The two variations of gap statistic, however, perform poorly in non-well-separated settings 6$\sim$10. On the other hand, silhouette performs better in settings 6$\sim$10 but not settings 6$\sim$10. S4 is the only method that generally performs among the top in all settings.
%&Predict&\multicolumn{3}{c|}{Stability}
\begin{table}	
	\caption{Clustering without feature selection simulation result: determining $K$ for $K$means. Every row indicates how many times this method chooses the correct $K$. Note that CH, KL and FW don't have mechanism to detect Null data.}\label{table:Lowdimensionsimulation}
	\hspace{-2cm}
	\begin{center}
		%\renewcommand\arraystretch{1.5}
		%\resizebox{\textwidth}{15mm}{
		\begin{tabular}{|c|c|c|c|c|c|c|c|c|c|c|c|}
			\hline
			&\multicolumn{7}{c|}{cluster tightness}&Predict&\multicolumn{3}{c|}{Stability}\\
			\cline{2-12}
			\diagbox{Setting}{method} & \rotatebox{90}{Gap/PCA}& \rotatebox{90}{Gap/Unif} & \rotatebox{90}{Jump}  & \rotatebox{90}{CH} & \rotatebox{90}{KL}  &\rotatebox{90}{H}&\rotatebox{90}{silhouette} &\rotatebox{90}{PS} &\rotatebox{90}{FW}&\rotatebox{90}{LD}&\rotatebox{90}{S4}\\
			\hline
			setting1   $K=1$            & 100 & 97 & 0  & - & -&-&-&100&-&88&98\\
			setting2   $K=3$           & 100  & 100 & 100 &  100 & 58&0&93&99&78&67&100 \\		
			setting3   $K=4$         & 99    & 100 & 99  & 98 & 71&2&76&98&68&53&99 \\	
			setting4    $K=4$         & 82    & 93 & 73  & 25 & 94&98&34&80&58&34&78\\
			setting5    $K=2$        &100     &0&0&0&100&0&100&65&100&100&91\\
			\hline
			setting6   $K=4$         &0      &0&21&21&27&0&30&0&17&31&40\\
			%&H index                 &--      &100&71&55&0\\
			setting7 $K=4$ & 2    & 5  & 53  & 49 & 38&0&66&1&56&56& 70 \\
			setting8 $K=4$ & 21   & 27   &  90  & 84 & 58 &0&93&15&94&68&79 \\
			setting9 $K=2$ & 1  & 0  &  0 & 0 & 25&0&100&67&62&78&87 \\
			setting10 $K=2$  & 0  & 0 &  0 & 0 & 0&0&0&15&0&16&4  \\
			\hline	
		\end{tabular}
	\end{center}
	
	%\begin{center}
		%\begin{tablenotes}
			
		%	\item*Methods included for extensive comparision in this paper.
			%\item[2] The last column indicates whether the methods are appropriate to be extended to estimate sparse parameter for sparse kmeans
		%\end{tablenotes}
	%\end{center}
\end{table}

\subsection{Simulation II: Simulations for Determining $K$ and $\lambda$ Simultaneously in Sparse $K$-Means with Independent Feature Structure}
\label{sec:Highdimionalsimulationindependent}

The following simulation is designed to evaluate methods (gap statistic, prediction strength and S4) for determining $K$ and $\lambda$ in sparse $K$-means when features are mutually independent. The implementation of gap statistic is based on the sparcl R package\citep{witten2013sparcl} and the reference data is generated by permutation in this package. We simulated three clusters, each with 33 subjects, and each subject has 1,000 features, of which $q$ features are informative to distinguish the three clusters while other features are random noises. Denote by $ X_{n\times p}$ the data matrix where $n=99$ and $p=1000$ and $x_{i,1:j}$ is the vector of subject $i$ with features from 1 to $j$. We simulate features by multivariate normal distribution. For the first $q$ predictive features, $x_{i,1:q} \sim mvrnorm(u, I_q)$ for $1 \leq i \leq 33$, $x_{i,1:q} \sim mvrnorm(0, I_q)$ for $34 \leq i \leq 66$, and $x_{i,1:q} \sim mvrnorm(-u, I_q)$ for $67 \leq i \leq 99$, where $u$ is the effect size to distinguish three clusters. For the remaining noise features, $x_{i,(q+1):p} \sim mvrnorm(0, I_{p-q})$ for $1\leq i\leq 99$. We choose $q=(50,200)$ and $u=(0.4, 0.6, 0.8)$ to generate six settings and repeat each setting for 50 times. We perform $B=100$ resampling evaluationfor all three methods and choose number of cluster $K$ from 2 to 7 for all six simulation settings.

Next, we evaluate the methods under two situations. Firstly, we assume $K=3$ is known and compare the performance of estimating $\lambda$, the sparsity parameter, and this is the same setting considered in \cite{witten2010framework}. Secondly, we consider simultaneous estimation of $K$ and $\lambda$. For both situations, we benchmark the clustering accuracy by adjusted Rand index(ARI)\citep{hubert1985comparing} when comparing to the underlying true clustering structure. We also benchmark feature selection by comparing selected features to the underlying true predictive features using Jaccard index\citep{jaccard1901distribution}, defined as $J(A,B)=|A\cap B|/|A\cup B|$ where $A$ is the set of selected features from sparse $K$-means and $B$ represents the set of $q$ true features. Table \ref{table:Highdimensionalsimulationindependent}A illustrates the result when $K=3$ is known and we apply extended Gap, PS and S4 to estimate $\lambda$. We observe that S4 universally achieves better clustering accuracy and more accurate feature selection in different $q$ and $u$. For example, when $u=0.8$ and $q=50$, S4 outperforms the other two methods with almost perfect clustering accuracy (average ARI=0.96 compared to 0.88 and 0.93 for Gap and PS) and feature selection (average Jaccard= 0.97 compared to 0.81 and 0.91 for Gap and PS). In the case of estimating $K$ and $\lambda$ simultaneously, Table \ref{table:Highdimensionalsimulationindependent}B shows ARI, Jaccard and the root mean square error (RMSE) of $K$ when selecting $K$ from 2 to 7. S4 again generally outperforms Gap and PS with average ARI=0.95 (compared to 0.47 for Gap and 0.85 for PS), average Jaccard=0.97 (compared to 0.66 for Gap and 0.89 for PS) and RMSE=0 (compared to 0.99 for Gap and 0.42 for PS) when $q=50$ and $u=0.8$. In general, PS performs slightly better than Gap but not as good as S4.
%(??did you use the same search rule as S4??).

% \usepackage{multirow}
% \usepackage{multirow}
% \usepackage{multirow}
% \usepackage{multirow}
% \usepackage{multirow}

\begin{table}
	\setlength\tabcolsep{2pt}
	\normalsize
	\begin{center}
		\caption{Simulation results for clustering with feature selection when features are mutually independent:table \ref{table:Highdimensionalsimulationindependent}A is the result for estimating $\lambda$ when $K$ is known; table \ref{table:Highdimensionalsimulationindependent}B is the result for simutaneous estimation of $K$ and $\lambda$. The value in each cell is the average index, and the value in the parenthesis is the standard deviation of 50 simulations.}\label{table:Highdimensionalsimulationindependent}
		\resizebox{\textwidth}{!}{
			\begin{tabular}{llllllll}
				\multicolumn{8}{c}{Estimation $\lambda$ when $K=3$ is known(Table \ref{table:Highdimensionalsimulationindependent}A)}
				\\ \hline
				\multicolumn{1}{|c|}{\multirow{3}{*}{Index}}                                                                   & \multicolumn{1}{l|}{\multirow{3}{*}{method}} & \multicolumn{3}{c|}{50 predictive genes}                                                                    & \multicolumn{3}{c|}{200 predictive genes}                                                                     \\ \cline{3-8} 
				\multicolumn{1}{|c|}{}                                                                                         & \multicolumn{1}{l|}{}                        & \multicolumn{3}{c|}{effect size}                                                                            & \multicolumn{3}{c|}{effect size}                                                                              \\ \cline{3-8} 
				\multicolumn{1}{|c|}{}                                                                                         & \multicolumn{1}{l|}{}                        & \multicolumn{1}{l|}{0.4}           & \multicolumn{1}{l|}{0.6}            & \multicolumn{1}{l|}{0.8}         & \multicolumn{1}{l|}{0.4}           & \multicolumn{1}{l|}{0.6}            & \multicolumn{1}{l|}{0.8}           \\ \hline
				\multicolumn{1}{|l|}{\multirow{3}{*}{\begin{tabular}[c]{@{}l@{}}Clustering\\ accuracy\\ (ARI)\end{tabular}}}   & \multicolumn{1}{l|}{Gap}                     & \multicolumn{1}{l|}{0.12(0.07)}    & \multicolumn{1}{l|}{0.46(0.14)}     & \multicolumn{1}{l|}{0.88(0.08)}  & \multicolumn{1}{l|}{0.69(0.21)}    & \multicolumn{1}{l|}{1(0.01)}        & \multicolumn{1}{l|}{1(0)}          \\ \cline{2-8} 
				\multicolumn{1}{|l|}{}                                                                                         & \multicolumn{1}{l|}{PS}                      & \multicolumn{1}{l|}{0.2(0.08)}     & \multicolumn{1}{l|}{0.57(0.14)}     & \multicolumn{1}{l|}{0.93(0.05)}  & \multicolumn{1}{l|}{0.8(0.09)}     & \multicolumn{1}{l|}{1(0.01)}        & \multicolumn{1}{l|}{1(0)}          \\ \cline{2-8} 
				\multicolumn{1}{|l|}{}                                                                                         & \multicolumn{1}{l|}{S4}                      & \multicolumn{1}{l|}{0.21(0.09)}    & \multicolumn{1}{l|}{0.65(0.13)}     & \multicolumn{1}{l|}{0.96(0.04)}  & \multicolumn{1}{l|}{0.79(0.08)}    & \multicolumn{1}{l|}{1(0.01)}        & \multicolumn{1}{l|}{1(0)}          \\ \hline
				\multicolumn{1}{|l|}{\multirow{3}{*}{\begin{tabular}[c]{@{}l@{}}Feature\\ selection\\ (Jaccard)\end{tabular}}} & \multicolumn{1}{l|}{Gap}                     & \multicolumn{1}{l|}{0.16(0.08)}    & \multicolumn{1}{l|}{0.6(0.11)}      & \multicolumn{1}{l|}{0.81(0.08)}  & \multicolumn{1}{l|}{0.53(0.15)}    & \multicolumn{1}{l|}{0.89(0.04)}     & \multicolumn{1}{l|}{0.94(0.03)}    \\ \cline{2-8} 
				\multicolumn{1}{|l|}{}                                                                                         & \multicolumn{1}{l|}{PS}                      & \multicolumn{1}{l|}{0.11(0.04)}    & \multicolumn{1}{l|}{0.2(0.23)}      & \multicolumn{1}{l|}{0.91(0.04)}  & \multicolumn{1}{l|}{0.36(0.07)}    & \multicolumn{1}{l|}{0.83(0.19)}     & \multicolumn{1}{l|}{0.95(0.02)}    \\ \cline{2-8} 
				\multicolumn{1}{|l|}{}                                                                                         & \multicolumn{1}{l|}{S4}                      & \multicolumn{1}{l|}{0.15(0.08)}    & \multicolumn{1}{l|}{0.79(0.1)}      & \multicolumn{1}{l|}{0.97(0.03)}  & \multicolumn{1}{l|}{0.6(0.06)}     & \multicolumn{1}{l|}{0.93(0.04)}     & \multicolumn{1}{l|}{0.99(0.02)}    \\ \hline
				\multicolumn{1}{|l|}{\multirow{3}{*}{\begin{tabular}[c]{@{}l@{}}Number \\ of feature\\ selected\end{tabular}}} & \multicolumn{1}{l|}{Gap}                     & \multicolumn{1}{l|}{63.48(27.24)}  & \multicolumn{1}{l|}{31.26(6.01)}    & \multicolumn{1}{l|}{40.48(3.9)}  & \multicolumn{1}{l|}{188(6)}        & \multicolumn{1}{l|}{179.52(9.21)}   & \multicolumn{1}{l|}{172.34(98.72)} \\ \cline{2-8} 
				\multicolumn{1}{|l|}{}                                                                                         & \multicolumn{1}{l|}{PS}                      & \multicolumn{1}{l|}{360(106.7)}    & \multicolumn{1}{l|}{466.08(220.26)} & \multicolumn{1}{l|}{45.6(2.1)}    & \multicolumn{1}{l|}{541.9(104.19)} & \multicolumn{1}{l|}{236.52(169.36)} & \multicolumn{1}{l|}{190.52(3.07)}  \\ \cline{2-8} 
				\multicolumn{1}{|l|}{}                                                                                         & \multicolumn{1}{l|}{S4}                      & \multicolumn{1}{l|}{246.18(98.16)} & \multicolumn{1}{l|}{48.4(12.16)}    & \multicolumn{1}{l|}{50.18(1.98)} & \multicolumn{1}{l|}{150.72(32.42)} & \multicolumn{1}{l|}{190.62(9.27)}   & \multicolumn{1}{l|}{198.3(3.15)}   \\ \hline    
				\\

				%\hline
				\multicolumn{8}{c}{Simultaneous estimation $\lambda$ and $K$(Table \ref{table:Highdimensionalsimulationindependent}B)}                                                                                                                                                         \\ \hline
				\multicolumn{1}{|c|}{\multirow{3}{*}{Index}}                                                                   & \multicolumn{1}{l|}{\multirow{3}{*}{method}} & \multicolumn{3}{c|}{50 predictive genes}                                                                     & \multicolumn{3}{c|}{200 predictive genes}                                                                   \\ \cline{3-8} 
				\multicolumn{1}{|c|}{}                                                                                         & \multicolumn{1}{l|}{}                        & \multicolumn{3}{c|}{effect size}                                                                             & \multicolumn{3}{c|}{effect size}                                                                            \\ \cline{3-8} 
				\multicolumn{1}{|c|}{}                                                                                         & \multicolumn{1}{l|}{}                        & 0.4                                 & 0.6                                 & \multicolumn{1}{l|}{0.8}         & 0.4                                & 0.6                                & \multicolumn{1}{l|}{0.8}          \\ \hline
				\multicolumn{1}{|l|}{\multirow{3}{*}{\begin{tabular}[c]{@{}l@{}}Clustering\\ accuracy\\ (ARI)\end{tabular}}}   & \multicolumn{1}{l|}{Gap}                     & \multicolumn{1}{l|}{0.12(0.07)}     & \multicolumn{1}{l|}{0.42(0.06)}     & \multicolumn{1}{l|}{0.47(0.08)}  & \multicolumn{1}{l|}{0.45(0.05)}    & \multicolumn{1}{l|}{0.55(0.07)}    & \multicolumn{1}{l|}{0.56(0.01)}   \\ \cline{2-8} 
				\multicolumn{1}{|l|}{}                                                                                         & \multicolumn{1}{l|}{PS}                      & \multicolumn{1}{l|}{0.16(0.1)}      & \multicolumn{1}{l|}{0.44(0.02)}     & \multicolumn{1}{l|}{0.85(0.19)}  & \multicolumn{1}{l|}{0.46(0.02)}    & \multicolumn{1}{l|}{0.98(0.09)}    & \multicolumn{1}{l|}{1(0)}         \\ \cline{2-8} 
				\multicolumn{1}{|l|}{}                                                                                         & \multicolumn{1}{l|}{S4}                      & \multicolumn{1}{l|}{0.26(0.11)}     & \multicolumn{1}{l|}{0.44(0.02)}     & \multicolumn{1}{l|}{0.95(0.04)}  & \multicolumn{1}{l|}{0.51(0.13)}    & \multicolumn{1}{l|}{1(0.01)}       & \multicolumn{1}{l|}{1(0)}         \\ \hline
				\multicolumn{1}{|l|}{\multirow{3}{*}{\begin{tabular}[c]{@{}l@{}}Feature\\ selection\\ (Jaccard)\end{tabular}}} & \multicolumn{1}{l|}{Gap}                     & \multicolumn{1}{l|}{0.16(0.08)}     & \multicolumn{1}{l|}{0.6(0.1)}       & \multicolumn{1}{l|}{0.66(0.09)}  & \multicolumn{1}{l|}{0.56(0.09)}    & \multicolumn{1}{l|}{0.74(0.08)}    & \multicolumn{1}{l|}{0.88(0.05)}   \\ \cline{2-8} 
				\multicolumn{1}{|l|}{}                                                                                         & \multicolumn{1}{l|}{PS}                      & \multicolumn{1}{l|}{0.08(0.02)}     & \multicolumn{1}{l|}{0.65(0.24)}     & \multicolumn{1}{l|}{0.89(0.06)}  & \multicolumn{1}{l|}{0.37(0.07)}    & \multicolumn{1}{l|}{0.7(0.29)}     & \multicolumn{1}{l|}{0.95(0.02)}   \\ \cline{2-8} 
				\multicolumn{1}{|l|}{}                                                                                         & \multicolumn{1}{l|}{S4}                      & \multicolumn{1}{l|}{0.22(0.12)}     & \multicolumn{1}{l|}{0.77(0.08)}     & \multicolumn{1}{l|}{0.97(0.03)}  & \multicolumn{1}{l|}{0.58(0.07)}    & \multicolumn{1}{l|}{0.92(0.04)}    & \multicolumn{1}{l|}{0.99(0.01)}   \\ \hline
				\multicolumn{1}{|l|}{\multirow{3}{*}{\begin{tabular}[c]{@{}l@{}}Number\\ of feature\\ selected\end{tabular}}}  & \multicolumn{1}{l|}{Gap}                     & \multicolumn{1}{l|}{75.54(49.89)}   & \multicolumn{1}{l|}{30.7(5.34)}     & \multicolumn{1}{l|}{32.92(4.7)}  & \multicolumn{1}{l|}{152.98(60.78)} & \multicolumn{1}{l|}{149.82(16.64)}  & \multicolumn{1}{l|}{175.66(10.87)} \\ \cline{2-8} 
				\multicolumn{1}{|l|}{}                                                                                         & \multicolumn{1}{l|}{PS}                      & \multicolumn{1}{l|}{543.34(111.86)} & \multicolumn{1}{l|}{116.78(118.33)} & \multicolumn{1}{l|}{44.4(3.11)}  & \multicolumn{1}{l|}{534.14(96.48)} & \multicolumn{1}{l|}{355.5(272.28)} & \multicolumn{1}{l|}{190.52(3.07)} \\ \cline{2-8} 
				\multicolumn{1}{|l|}{}                                                                                         & \multicolumn{1}{l|}{S4}                      & \multicolumn{1}{l|}{200.42(126.68)} & \multicolumn{1}{l|}{42.56(6.13)}    & \multicolumn{1}{l|}{49.92(1.88)} & \multicolumn{1}{l|}{145.32(32.18)} & \multicolumn{1}{l|}{189.58(9.53)}  & \multicolumn{1}{l|}{198.36(2.99)} \\ \hline
				\multicolumn{1}{|l|}{\multirow{3}{*}{\begin{tabular}[c]{@{}l@{}}RMSE \\ of K\end{tabular}}}                    & \multicolumn{1}{l|}{Gap}                     & \multicolumn{1}{l|}{1.44}           & \multicolumn{1}{l|}{0.99}           & \multicolumn{1}{l|}{0.99}        & \multicolumn{1}{l|}{1}             & \multicolumn{1}{l|}{0.99}          & \multicolumn{1}{l|}{1}            \\ \cline{2-8} 
				\multicolumn{1}{|l|}{}                                                                                         & \multicolumn{1}{l|}{PS}                      & \multicolumn{1}{l|}{2.34}           & \multicolumn{1}{l|}{1}              & \multicolumn{1}{l|}{0.42}        & \multicolumn{1}{l|}{1}             & \multicolumn{1}{l|}{0.2}           & \multicolumn{1}{l|}{0}            \\ \cline{2-8} 
				\multicolumn{1}{|l|}{}                                                                                         & \multicolumn{1}{l|}{S4}                      & \multicolumn{1}{l|}{1.62}           & \multicolumn{1}{l|}{1}              & \multicolumn{1}{l|}{0}           & \multicolumn{1}{l|}{0.94}          & \multicolumn{1}{l|}{0}             & \multicolumn{1}{l|}{0}            \\ \hline
				%			&                                              &            &            &                                 &            &            &                                 \\
				%			&                                              &            &            &                                 &            &            &                                 \\
				%			&                                              &            &            &                                 &            &            &                                 \\
				%			&                                              &            &            &                                 &            &            &                                 \\
				%			&                                              &            &            &                                 &            &            &                                 \\
				%			&                                              &            &            &                                 &            &            &                                 \\
				&                                              &            &            &                                 &            &            &                                  
		\end{tabular}}
		%\begin{itemize}
		%	\item The value in the parentheses is standard deviation all the result of simulations\\
	%	\end{itemize}

	\end{center}
	%\begin{center}
	%\hspace{-5cm}
	%	\vspace{-\topsep}
	%	\begin{tablenotes}
	
	%		\item "Prok" stands for the proportion of times to choose the correct number of cluster $K$ in 50 simulations
	%\item[2] The last column indicates whether the methods are appropriate to be extended to estimate sparse parameter for sparse kmeans
	%	\end{tablenotes}
	%\end{center}
	%
\end{table}

\subsection{Simulation III: Simulations for Determining $K$ and $\lambda$ Simultaneously in Sparse $K$-Means with Correlated Feature Structure}
To better mimic the nature of gene expression profile data from microarray or RNA-seq experiments, a typical high-dimensional data type for clustering, we simulate data of three clusters with gene correlation structure as co-regulated gene modules. The purpose is usually to cluster patients to identify novel disease subtypes. Below are the detailed steps to simulate clustered genes and noise genes. 

\textit{\underline{Simulation of cluster predictive genes:}} 
\begin{itemize*}
	\item[1.]Simulate number of subjects $N_1$, $N_2$ and $N_3$ for three disease subtypes by sampling from Poisson distribution with mean 40 30 and 20 respectively. The total number of subjects in each simulated data is $N=N_1+N_2+N_3$
	\item[2.]Simulate $M$ gene modules. In each module, sample $n_m(1\leq m\leq M)$ genes from $POI(20)$. Therefore, there will be an average of $20\times M$ predictive genes to characterize the three clusters (disease subtypes).
	\item[3.]Simulate $u_{km}\sim U(4, 10)$ with constrain $U_{lower} \leq max_{p,q}|u_{pm}-u_{qm}| \leq U_{upper}$, where $u_{km}$ is the template gene expression of cluster $k$ $(1 \leq k \leq 3)$ and module $m$ $(1\leq m \leq M)$ and $(U_{lower},U_{upper})$ is effect size.
	\item[4.]Add biological variation $\sigma_1^2$ to the template gene expression and simulate $X_{kmi}\sim N(u_{km},\sigma_1^2)$ for each module $m$, subject $i$ ($1\leq i \leq N_k$) of cluster $k$.
	\item[5.]Simulate covariance matrix $\Sigma_{mk}$ for genes in module $m$ ($1\leq m \leq M$) and cluster $k$ ($1\leq k \leq 3$). First simulate $\Sigma^{'}_{mk}$ from inverse Wishart distribution, $W^{-1}(\Phi,60)$ where $\Phi=(1-\phi_{cov})\cdot I_{n_m\times n_m}+\phi_{cov}\cdot J_{n_m\times n_m}$, $I$ is identity matrix, $J$ is a matrix with all elements equivalent to 1 and $\phi_{cov}$ is a scalar controlling degree of correlation among genes, Then $\Sigma_{mk}$ is calculated by standardizing $\Sigma^{'}_{mk}$ such that the diagonal elements are all 1.
	\item[6.]Simulate gene expression levels of genes in module $m$ for sample $i$ in cluster $k$ as $(X_{1kmi}, \cdots, X_{n_mkmi})$ from multivariate normal distribution with mean $X_{kmi}$ and covariance matrix $ \Sigma_{mk}$, where $1\leq i\leq N_k$, $1 \leq m \leq M$, $1\leq k \leq 3$.
\end{itemize*}

\textit{\underline{Simulation of noise genes:}} 
\begin{itemize}
	\item[1.]Simulate 600 noise genes.  For each gene, we generate the mean template gene expression $u_g \sim U(4, 10)$, where $1\leq g \leq 600$. 
	\item[2.]Then we add biological variation variance $\sigma_2^2$ to simulate gene expression level $X_{gi} \sim N(u_g, \sigma^2_2)$, $1 \leq i \leq N$. 
\end{itemize}
Gene expression levels of noise genes are relatively stable. Therefore, these genes could be regarded as housekeeping genes if their expression levels are high, or non-expressed genes if their expression levels are low.

We fix $\sigma_1^2=0.2, \sigma_2^2=1, M=10$ while tuning effect size $(U_{upper},U_{lower})$ and correlation parameter $\phi_{cov}$ to compare S4 with Gap and PS in different scenarios. Since the number of predictive genes in each gene module follows $POI(20)$, so the average number of predictive genes in each dataset is 200. Number of clusters $K$ is selected from 2 to 7 and ARI, Jaccard and RMSE of $K$ are used to compare the performance. Each setting is repeated 50 times. 

Table \ref{table:Sparsecov}A summarizes the result when $K=3$ is known and we apply Gap, PS and S4 to estimate $\lambda$. When $K=3$ is known, all three methods have high clustering accuracy (ARI) with PS and S4 as the better performers. S4 outperforms universally better in feature selection in different level of feature correlation $\phi_{cov}$ and effect size. Particularly, when $\phi_{cov}=0.1$ and ($U_{lower}$, $U_upper$)=(0.8, 1.0), the average Jaccard index for S4 is 0.72 compared to 0.6 for Gap and 0.44 for PS. Table \ref{table:Sparsecov}B shows ARI, Jaccard and RMSE when estimating $K$ and $\lambda$ simultaneously. S4 generally outperforms Gap and PS in cluster accuracy in terms of average ARI and feature selection in terms of average Jaccard index. S4 also has better performance in estimating $K$ with much smaller RMSE.

% \usepackage{multirow}
% \usepackage{multirow}
% \usepackage{multirow}
% \usepackage{multirow}

\begin{table}[]
	\setlength\tabcolsep{2pt}
	\begin{center}
		\caption{Simulation result for clustering with feature selection when features are correlated: table \ref{table:Sparsecov}A is the result for estimating $\lambda$ when $K$ is known; table \ref{table:Sparsecov}B is the result for simutaneous estimation of $K$ and $\lambda$. The value in each cell is the average index, and the value in the parenthesis is the standard deviation of 50 simulations.}\label{table:Sparsecov}
		% \usepackage{multirow}
		% \usepackage{multirow}
		\resizebox{\textwidth}{!}{
			\begin{tabular}{ccllllll}
				\multicolumn{8}{c}{Estimation $\lambda$ when K=3 is known (Table \ref{table:Sparsecov}A) }                                                                                                                                                                                                                        \\
				 \hline
				\multicolumn{1}{|c|}{\multirow{3}{*}{Index}}                                                                   & \multicolumn{1}{c|}{\multirow{3}{*}{method}} & \multicolumn{2}{c|}{$\phi_{cov}$=0.1}                                             & \multicolumn{2}{c|}{$\phi_{cov}$=0.3}                                            & \multicolumn{2}{c|}{$\phi_{cov}$=0.5}                                            \\ \cline{3-8} 
				\multicolumn{1}{|c|}{}                                                                                         & \multicolumn{1}{c|}{}                        & \multicolumn{2}{c|}{effect size: ($U_{lower}, U_{upper}$)}                                       & \multicolumn{2}{c|}{effect size: ($U_{lower}, U_{upper}$)}                                        & \multicolumn{2}{c|}{effect size: ($U_{lower}, U_{upper}$)}                                        \\ \cline{3-8} 
				\multicolumn{1}{|c|}{}                                                                                         & \multicolumn{1}{c|}{}                        & \multicolumn{1}{l|}{(0.8, 1.0)}    & \multicolumn{1}{l|}{(1.0, 1.5)}     & \multicolumn{1}{l|}{(1.5, 2)}      & \multicolumn{1}{l|}{(2, 2.5)}      & \multicolumn{1}{l|}{(1.5, 2)}      & \multicolumn{1}{l|}{(2, 2.5)}      \\ \hline
				\multicolumn{1}{|c|}{\multirow{3}{*}{\begin{tabular}[c]{@{}c@{}}Cluster \\ accuracy\\ (ARI)\end{tabular}}}     & \multicolumn{1}{c|}{Gap}                     & \multicolumn{1}{l|}{0.84(0.14)}    & \multicolumn{1}{l|}{0.97(0.09)}     & \multicolumn{1}{l|}{0.99(0.02)}     & \multicolumn{1}{l|}{1(0.01)}       & \multicolumn{1}{l|}{0.8(0.22)}     & \multicolumn{1}{l|}{0.97(0.08)}    \\ \cline{2-8} 
				\multicolumn{1}{|c|}{}                                                                                         & \multicolumn{1}{c|}{PS}                      & \multicolumn{1}{l|}{0.89(0.08)}    & \multicolumn{1}{l|}{0.98(0.03)}     & \multicolumn{1}{l|}{0.98(0.04)}    & \multicolumn{1}{l|}{1(0.01)}       & \multicolumn{1}{l|}{0.92(0.12)}    & \multicolumn{1}{l|}{0.99(0.04)}    \\ \cline{2-8} 
				\multicolumn{1}{|c|}{}                                                                                         & \multicolumn{1}{c|}{S4}                      & \multicolumn{1}{l|}{0.88(0.09)}    & \multicolumn{1}{l|}{0.98(0.05)}     & \multicolumn{1}{l|}{0.97(0.1)}     & \multicolumn{1}{l|}{1(0)}          & \multicolumn{1}{l|}{0.9(0.16)}     & \multicolumn{1}{l|}{0.98(0.08)}    \\ \hline
				\multicolumn{1}{|c|}{\multirow{3}{*}{\begin{tabular}[c]{@{}c@{}}Feature\\ selection\\ (Jaccard)\end{tabular}}} & \multicolumn{1}{c|}{Gap}                     & \multicolumn{1}{l|}{0.6(0.19)}     & \multicolumn{1}{l|}{0.79(0.15)}     & \multicolumn{1}{l|}{0.8(0.23)}     & \multicolumn{1}{l|}{0.94(0.06)}    & \multicolumn{1}{l|}{0.65(0.3)}     & \multicolumn{1}{l|}{0.87(0.21)}    \\ \cline{2-8} 
				\multicolumn{1}{|c|}{}                                                                                         & \multicolumn{1}{c|}{PS}                      & \multicolumn{1}{l|}{0.44(0.11)}    & \multicolumn{1}{l|}{0.78(0.19)}     & \multicolumn{1}{l|}{0.87(0.15)}    & \multicolumn{1}{l|}{0.97(0.03)}    & \multicolumn{1}{l|}{0.83(0.17)}    & \multicolumn{1}{l|}{0.93(0.14)}    \\ \cline{2-8} 
				\multicolumn{1}{|c|}{}                                                                                         & \multicolumn{1}{c|}{S4}                      & \multicolumn{1}{l|}{0.72(0.08)}    & \multicolumn{1}{l|}{0.88(0.09)}     & \multicolumn{1}{l|}{0.92(0.12)}     & \multicolumn{1}{l|}{0.99(0.02)}    & \multicolumn{1}{l|}{0.81(0.22)}    & \multicolumn{1}{l|}{0.92(0.18)}    \\ \hline
				\multicolumn{1}{|c|}{\multirow{3}{*}{\begin{tabular}[c]{@{}c@{}}Number\\ of feature\\ selected\end{tabular}}}  & \multicolumn{1}{c|}{Gap}                     & \multicolumn{1}{l|}{164.04(86.85)} & \multicolumn{1}{l|}{165.62(39.87)}  & \multicolumn{1}{l|}{161.74(47.76)} & \multicolumn{1}{l|}{189.52(19.71)} & \multicolumn{1}{l|}{130.62(59.77)} & \multicolumn{1}{l|}{174.88(45.32)} \\ \cline{2-8} 
				\multicolumn{1}{|c|}{}                                                                                         & \multicolumn{1}{c|}{PS}                      & \multicolumn{1}{l|}{469.84(90.52)} & \multicolumn{1}{l|}{237.14(133.96)} & \multicolumn{1}{l|}{173.62(31.84)} & \multicolumn{1}{l|}{195.38(15.08)} & \multicolumn{1}{l|}{166.76(36.19)} & \multicolumn{1}{l|}{186.64(32.68)} \\ \cline{2-8} 
				\multicolumn{1}{|c|}{}                                                                                         & \multicolumn{1}{c|}{S4}                      & \multicolumn{1}{l|}{169.18(27.3)}  & \multicolumn{1}{l|}{180.16(24.3)}   & \multicolumn{1}{l|}{178.04(42.77)} & \multicolumn{1}{l|}{198.86(15.13)} & \multicolumn{1}{l|}{162.7(44.19)}  & \multicolumn{1}{l|}{185.86(39.94)} \\ \hline   
				\\
				\multicolumn{8}{c}{Simultaneous estimation $\lambda$ and $K$ (Table \ref{table:Sparsecov}B)}                                                                                                                                                                                                                         \\ \hline
				\multicolumn{1}{|c|}{\multirow{3}{*}{Index}}                                                                   & \multicolumn{1}{c|}{\multirow{3}{*}{method}} & \multicolumn{2}{c|}{$\phi_{cov}$=0.1}                                             & \multicolumn{2}{c|}{$\phi_{cov}$=0.3}                                            & \multicolumn{2}{c|}{$\phi_{cov}$=0.5}                                            \\ \cline{3-8} 
				\multicolumn{1}{|c|}{}                                                                                         & \multicolumn{1}{c|}{}                        & \multicolumn{2}{c|}{effect size: ($U_{lower}, U_{upper}$)}                                       & \multicolumn{2}{c|}{effect size: ($U_{lower}, U_{upper}$)}                                        & \multicolumn{2}{c|}{effect size: ($U_{lower}, U_{upper}$)}                                        \\ \cline{3-8} 
				\multicolumn{1}{|c|}{}                                                                                         & \multicolumn{1}{c|}{}                        & \multicolumn{1}{l|}{(0.8, 1.0)}     & \multicolumn{1}{l|}{(1.0, 1.5)}    & \multicolumn{1}{l|}{(1.5, 2)}      & \multicolumn{1}{l|}{(2, 2.5)}      & \multicolumn{1}{l|}{(1.5, 2)}      & \multicolumn{1}{l|}{(2, 2.5)}      \\ \hline
				\multicolumn{1}{|c|}{\multirow{3}{*}{\begin{tabular}[c]{@{}c@{}}Cluster\\ accuracy\\ (ARI)\end{tabular}}}      & \multicolumn{1}{l|}{Gap}                     & \multicolumn{1}{l|}{0.64(0.21)}     & \multicolumn{1}{l|}{0.74(0.15)}    & \multicolumn{1}{l|}{0.84(0.16)}    & \multicolumn{1}{l|}{0.84(0.17)}    & \multicolumn{1}{l|}{0.72(0.19)}    & \multicolumn{1}{l|}{0.8(0.18)}     \\ \cline{2-8} 
				\multicolumn{1}{|c|}{}                                                                                         & \multicolumn{1}{l|}{PS}                      & \multicolumn{1}{l|}{0.78(0.15)}     & \multicolumn{1}{l|}{0.89(0.16)}    & \multicolumn{1}{l|}{0.95(0.1)}     & \multicolumn{1}{l|}{0.97(0.11)}    & \multicolumn{1}{l|}{0.83(0.17)}    & \multicolumn{1}{l|}{0.91(0.16)}    \\ \cline{2-8} 
				\multicolumn{1}{|c|}{}                                                                                         & \multicolumn{1}{l|}{S4}                      & \multicolumn{1}{l|}{0.77(0.16)}     & \multicolumn{1}{l|}{0.97(0.07)}    & \multicolumn{1}{l|}{0.97(0.08)}    & \multicolumn{1}{l|}{1(0)}          & \multicolumn{1}{l|}{0.87(0.15)}    & \multicolumn{1}{l|}{0.98(0.07)}    \\ \hline
				\multicolumn{1}{|c|}{\multirow{3}{*}{\begin{tabular}[c]{@{}c@{}}Feature\\ selection\\ (Jaccard)\end{tabular}}} & \multicolumn{1}{l|}{Gap}                     & \multicolumn{1}{l|}{0.5(0.19)}      & \multicolumn{1}{l|}{0.56(0.18)}    & \multicolumn{1}{l|}{0.71(0.23)}    & \multicolumn{1}{l|}{0.74(0.25)}    & \multicolumn{1}{l|}{0.62(0.25)}    & \multicolumn{1}{l|}{0.7(0.25)}     \\ \cline{2-8} 
				\multicolumn{1}{|c|}{}                                                                                         & \multicolumn{1}{l|}{PS}                      & \multicolumn{1}{l|}{0.4(0.1)}       & \multicolumn{1}{l|}{0.66(0.23)}    & \multicolumn{1}{l|}{0.89(0.11)}    & \multicolumn{1}{l|}{0.95(0.1)}     & \multicolumn{1}{l|}{0.79(0.18)}    & \multicolumn{1}{l|}{0.9(0.18)}     \\ \cline{2-8} 
				\multicolumn{1}{|c|}{}                                                                                         & \multicolumn{1}{l|}{S4}                      & \multicolumn{1}{l|}{0.65(0.13)}     & \multicolumn{1}{l|}{0.87(0.1)}     & \multicolumn{1}{l|}{0.92(0.12)}    & \multicolumn{1}{l|}{0.99(0.02)}    & \multicolumn{1}{l|}{0.81(0.23)}    & \multicolumn{1}{l|}{0.93(0.17)}    \\ \hline
				\multicolumn{1}{|c|}{\multirow{3}{*}{\begin{tabular}[c]{@{}c@{}}Number\\ of feature\\ selected\end{tabular}}}  & \multicolumn{1}{l|}{Gap}                     & \multicolumn{1}{l|}{114.18(67.69)}  & \multicolumn{1}{l|}{113.72(34.83)} & \multicolumn{1}{l|}{141.66(46.67)} & \multicolumn{1}{l|}{149.68(52.51)} & \multicolumn{1}{l|}{124.66(48.84)} & \multicolumn{1}{l|}{142.32(51.9)}  \\ \cline{2-8} 
				\multicolumn{1}{|c|}{}                                                                                         & \multicolumn{1}{l|}{PS}                      & \multicolumn{1}{l|}{485.04(102.23)} & \multicolumn{1}{l|}{275.3(195.23)} & \multicolumn{1}{l|}{178.88(26.69)} & \multicolumn{1}{l|}{190.92(24.97)} & \multicolumn{1}{l|}{168.68(66.46)} & \multicolumn{1}{l|}{180.54(39.9)}  \\ \cline{2-8} 
				\multicolumn{1}{|c|}{}                                                                                         & \multicolumn{1}{l|}{S4}                      & \multicolumn{1}{l|}{151.02(35.66)}  & \multicolumn{1}{l|}{179.3(25.1)}   & \multicolumn{1}{l|}{185.44(27.28)} & \multicolumn{1}{l|}{198.94(15.2)}  & \multicolumn{1}{l|}{161.8(46.08)}  & \multicolumn{1}{l|}{186.98(38.99)} \\ \hline
				\multicolumn{1}{|l|}{\multirow{3}{*}{\begin{tabular}[c]{@{}l@{}}RMSE\\ of K\end{tabular}}}                     & \multicolumn{1}{l|}{Gap}                     & \multicolumn{1}{l|}{0.87}           & \multicolumn{1}{l|}{0.88}          & \multicolumn{1}{l|}{0.71}          & \multicolumn{1}{l|}{0.71}          & \multicolumn{1}{l|}{0.81}          & \multicolumn{1}{l|}{0.76}          \\ \cline{2-8} 
				\multicolumn{1}{|l|}{}                                                                                         & \multicolumn{1}{l|}{PS}                      & \multicolumn{1}{l|}{0.69}           & \multicolumn{1}{l|}{0.55}          & \multicolumn{1}{l|}{0.37}          & \multicolumn{1}{l|}{0.28}          & \multicolumn{1}{l|}{0.62}          & \multicolumn{1}{l|}{0.47}          \\ \cline{2-8} 
				\multicolumn{1}{|l|}{}                                                                                         & \multicolumn{1}{l|}{S4}                      & \multicolumn{1}{l|}{0.71}           & \multicolumn{1}{l|}{0.24}          & \multicolumn{1}{l|}{0.24}          & \multicolumn{1}{l|}{0}             & \multicolumn{1}{l|}{0.53}          & \multicolumn{1}{l|}{0.2}           \\ \hline

		\end{tabular}}
		
	\end{center}
\end{table}

\section{Real Applications}
\label{sec:Application}
We select a variety of datasets including four microarray datasets, two RNA sequencing datasets, one single nucleotide polymorphism (SNP) dataset and two non-omics datasets to evaluate the performance of S4 method. Table \ref{table:realdata} outlines the datasets and more details are presented in Section 6.1.

%We apply S4 to three Leukemia transcriptomic studies: \citep{verhaak2009prediction}, \citep{balgobind2010evaluation} and \citep{kohlmann2008international}. For every study we only consider samples from Acute myeloid leukima and there will be three clusters in each study where every sample will be subtype inv(16)(inversions in chromosome16), t(15;17)(translocations between chromosome 15 and 17), or t(8;21)(translocations between chromosome 8 and 21). We treat all the class label as underlying truth and apply S4, gap statistic, and prediction strength respectively to each study. The clustering performance is measured by ARI and whether the method choose the correct $K$. The feature selection performance is measured by looking at patterns in the heatmap. All the datasets were downloaded directly from NCBI GEO website. Originally there are 54676 genes in each study. We remove the probes with missing values and select the probe with largest interquartile region to represent the gene if multiple probes are mapped to the same gene. There are 20192 genes for every study after this preprocessing. Furthermore, For every study, we take the data in log scale and only take the top 10000 genes with largest mean value, which is aimed to filter genes with low expression level. Then we take top 5000 genes with largest variance. Finally, there are three datasets, each with 5000 genes. The summary of three preprocessed datasets is in Table 5.
%
\newpage
\begin{table}
	\center
	\caption {Summary of all the datasets after preprocessing}\label{table:realdata} 
	\setlength\tabcolsep{2pt}
	\hspace{-2cm}
	\resizebox{\textwidth}{!}{
		\begin{tabular}{|c|c|c|c|c|c|c|c|}
			\hline
			Data Type&Data description&source	&  \tabincell{c}{Number \\of features \\used} &\tabincell{c}{Number \\of \\samples} & True class label \\
			\hline
			\multirow{4}*{Microarray}&Leukemia&	\cite{verhaak2009prediction}& 2000 &89 & (33, 21, 35)\\
			&Leukemia&	\cite{balgobind2010evaluation}& 2000  & 74 & (27, 19, 28)\\
			&Leukemia&	\cite{kohlmann2008international} & 2000 & 105  & (28, 37, 40)\\
			&Mammalian tissue&	\cite{su2002large}&2000&102& (25, 26, 28 ,23)\\
			\hline
			\multirow{2}*{RNA sequencing}&Rat brain&	\cite{li2013transcriptome}&2000&36& (12,12,12)\\
			&Pan-cancer&	UCI repository&2000&801& (300, 146, 78 ,141,136)\\
			\hline
			SNP&SNP&	HapMap Consortium&17026&293& (71,151,71)\\
			\hline
			\multirow{2}*{Non-Omics}&Plant leaves&	\cite{mallah2013plant}&190&64& (16,16,16,16)\\
			&ISOLET&	UCI repository&617&1200& (240,240,240,240,240)\\
			\hline	
	\end{tabular}}
	%\begin{tablenotes}
	%\footnotesize Br Pr Lu Co
	%\item[1] *Affymetrix human genome u133 plus 2.0 array	
	\small
	\begin{itemize*}
		\item websites for the datasets in order
		\item[1] \url{https://www.ncbi.nlm.nih.gov/geo/query/acc.cgi?acc=GSE6891}
		\item[2] \url{https://www.ncbi.nlm.nih.gov/geo/query/acc.cgi?acc=GSE17855}\\
		\item[3] \url{https://www.ncbi.nlm.nih.gov/geo/query/acc.cgi?acc=GSE13159}\\
		\item[4] \url{http://portals.broadinstitute.org/cgi-bin/cancer/datasets.cgi}\\
		\item[5] \url{https://www.ncbi.nlm.nih.gov/geo/query/acc.cgi?acc=GSE47474}\\
		\item[6] \url{https://archive.ics.uci.edu/ml/datasets/gene+expression+cancer+RNA-Seq}\\
		\item[7] \url{ftp://ftp.ncbi.nlm.nih.gov/hapmap/genotypes/2008-07_phaseIII/hapmap_format/forward/}
		\item[8] \url{https://archive.ics.uci.edu/ml/datasets/One-hundred+plant+species+leaves+data+set}
		\item[9] \url{https://archive.ics.uci.edu/ml/datasets/isolet}
	\end{itemize*}
	
	%	\item[2] *The class label for \cite{su2002large} means breast, prostate, lung, and colon.
	%\end{tablenotes}
\end{table}
\subsection{Data description}
6.1.1 \quad Microarray  datasets\\
\textit{\underline{Three leukemia datasets:}} We collect three leukemia transcriptomic studies for evaluation: \cite{verhaak2009prediction}, \cite{balgobind2010evaluation} and \cite{kohlmann2008international}. For every study we only considered samples from acute myeloid leukemia with three pre-detected chromosome translocation subtypes: inv(16)(inversions in chromosome 16), t(15;17)(translocations between chromosomes 15 and 17), or t(8;21)(translocations between chromosomes 8 and 21). All the datasets are downloaded directly from NCBI GEO website with  GSE6891\citep{verhaak2009prediction}, GSE17855\citep{balgobind2010evaluation} and GSE13159\citep{kohlmann2008international}. Originally there are 54,676 probesets in each dataset and we remove the probesets with missing values and select the probesets with the largest interquartile range to represent the gene if multiple probesets are mapped to the same gene. 20,192 unique genes remained for every study after this preprocessing. Furthermore, for each study, we transform data to log scale and only keep the top 10,000 genes with the largest mean expression level (i.e. filter out low-expressed genes). We further filter out 8,000 genes with smaller variance (i.e. genes with little predictive information). Finally, the remaining 2,000 genes are used in the analysis.
%We treated the class label as underlying truth and applied Gap, PS and S4 to each study. The clustering performance is measured by ARI and %whether the method chooses the correct $K$. Since the true predictive features are not known, the feature selection performance cannot be %judged and we only display the expression patterns of selected features in the heatmap. 

\textit{\underline{Mammalian tissue types dataset:}} Gene expression from human and mouse samples across a diverse array of tissues, organs,
and cell lines have been profiled by \cite{su2002large}. Here we only consider four tissue types: breast, prostate, lung, and colon, which is available in R package fabiaData (Hochreiter et al., 2013) and website \url{http://portals.broadinstitute.org/cgi-bin/cancer/datasets.cgi}. The original data has 102 samples and 5565 probesets (genes). Following similar preprocessing procedure, We keep 3000 genes with the highest mean expression value and then 2000 genes were used in the analysis after further filtering low-variance genes.

%verhaak 6891
%balgobind 17855

%Table 6 shows estimation results of Gap, PS and S4 applied on the three examples. Gap and S4 captured the correct number of clusters ($K=3$) %n all three studies while PS estimate $K=2$ in \cite{balgobind2010evaluation}. Gap selected $\lambda$ value that correspond almost all genes in all %three examples, while S4 selected more reasonable number of genes: 756, 90 and 74. With less selected genes than Gap, S4 achieved better ARI %(0.93, 0.96 and 0.97 compared to Gap’s 0.89, 0.83 and 0.94), which argues that S4 selected fewer but possibly more accurate genes to achieve %better clustering. In fact, heatmaps of Gap and S4 results in Figure 2 visually confirms the better clustering performance of S4. PS selected few %genes and  achieved the worse ARI in \cite{kohlmann2008international}(0.95) and \cite{balgobind2010evaluation}(0.48). For %\cite{verhaak2009prediction}, PS achieves best ARI 0.97. From this real example with biological underlying truth of clusters (but no feature %selection truth), S4 outperforms Gap and PS in estimating $K$ and $\lambda$.

6.1.2 \quad RNA sequencing data\\
\textit{\underline{Multiple brain regions of rat dataset:}} \cite{li2013transcriptome} generated a rat experiment including multiple brain regions (GSE47474) . RNA samples from three brain regions (hippocampus, striatum and prefrontal cortex) were sequenced for both control strains and HIV strains. Only the 36 control strains (12 in each brain region) are used here to see whether samples from three brain regions can be correctly clustered ($K$ = 3; $n_1$ = $n_2$ = $n_3$ $= 12$). The original count data is transformed into CPM value followed by log transformation and then 2000 genes are kept by filtering low-expressed genes and low-variance genes.

\noindent \textit{\underline{Pan-cancer dataset:}} We download a dataset which is part of the cancer genome atlas pan-cancer analysis project, available at UCI machine learning repository (\url{https://archive.ics.uci.edu/ml/datasets/gene+expression+cancer+RNA-Seq#} ). This collection of data consits of  five different types of tumor: 300 breast cancer (BRCA), 146 kidney clear cell carcinoma (KIRC), 78 colon cancer (COAD), 141 lung adenocarcinoma (LUAD) and 136 prostate cancer (PRAD). The data has already been normalized and we use the same filtering process to keep 2000 genes.\\
6.1.3 \quad SNP dataset\\
The SNP data is the one presented in \cite{witten2010framework}, where they showed that when number of cluster is known as three, the gap statistic will seemingly overestimate the number of features with non-zero weight. The data is publicly-available from Haplotype Map (“HapMap”) data of the International HapMap Consortium. Following the same preprocessing procedure as \cite{witten2010framework}, only phase III SNP data is used and we restrict the analysis to chromosome 22 of three populations: African ancestry in southwest USA (ASW), Utah residents with European ancestry (CEU), and Han Chinese from Beijing (CHB) since these three populations are known to be genetically distinct. All the available SNPs on chromesome22 are considered in the data, which gives us 293 samples and 17026 SNP. We then coded AA as 2, Aa as 1 and aa as 0, and use 5-nearest neighbors method \citep{troyanskaya2001missing} to impute the missing data.
 
6.1.4 \quad Non-Omics data\\
\noindent \textit{\underline{Plant species leaves dataset :}} \cite{mallah2013plant} introduced a dataset consisting of one-hundred species of plants with three types of features for leaves: shape texture and margin. Here we only consider 4 species out of 100, Acer Mono, Acer Palmatum, Acer Pictum and Acer Capillipes. After deleting features with any missing values, we have 64 samples (16 for each species) and 187 features.

\noindent \textit{\underline{ISOLET Data Set:}} ISOLET dataset was generated by a study where 150 subjects spoke each letter of the alphabet twice and recorded 617 features including spectral coefficients, contour features, sonorant features, pre-sonorant features and post-sonorant features. We only use five vowels and 1200 training subjects ( 240 samples for each of five vowels). Both plant species dataset and ISOLET dataset are publicly available in the UCI machine learning repository \citep{Dua:2017}.
\subsection{Results} 
Table \ref{table:Realdataresult} summarizes the results from three methods for all aforementioned real datasets. S4 outperforms Gap and PS in almost all applications, indicating that S4 is a generally robust framework to simultaneouly estimate $K$ and $\lambda$ in these Omics and non-Omics studies. The details are discussed below.

\noindent 6.2.1 \quad Microarray  datasets\\
For these three leukemia datasets, Gap and S4 captures the correct number of clusters ($K=3$) in all three studies while PS estimate $K=2$ for the Kohlmann study and Verhaak study. Gap selects 625, 491 and 330 genes in three datasets respectively, while S4 selects more manageable number of genes: 24, 64 and 81 respectively. Although S4 selects fewer number of genes than Gap, it even achieves slightly better ARI (0.94, 0.96 and 0.97 compared to Gaps 0.89, 0.96 and 0.94), which implies that S4 selected fewer but possibly more predictive genes to achieve better clustering. From these three real examples with known biological underlying clusters (but no truth for feature selection), S4 outperforms Gap and PS in estimating $K$ and $\lambda$. For mammalian tissue types data \citep{su2002large}, although  all three methods fail to select the correct $K=4$, S4 method estimates $K=3$ and achieves better ARI(0.65) than that of PS (0.33) and gap statistic (0.33), which both estimate $K=2$. Again, S4 determines a smaller number of features but achieves a better ARI.

\noindent 6.2.2 \quad RNA sequencing data\\
For rat brain data \citep{li2013transcriptome}, all three methods correctly estimate $K$ and cluster samples perfectly with $ARI=1$, whereas S4 selects smaller number of genes (79) than PS (1584) and gap statistic (295). This again suggests S4 potentially selects a group of genes that are informative enough. For Pan-cancer data, although all three methods fail to select $K=5$, S4 method has the closest result with $K=6$ and $ARI=0.78$.

\noindent 6.2.3 \quad SNP dataset\\
For SNP data, all three methods correctly identify $K=3$. \cite{witten2010framework} selected 7160 SNPs with non-zero weights using one standard deviation rule and mentioned the need for a more accurate method for tuning parameter selection. From our result, Gap statistic selects 7397 SNPs, producing similar number as \cite{witten2010framework}, whereas S4 selects 5595 SNPs and achieves exactly the same ARI as Gap statistic (0.92), suggesting better feature selection by S4.

\noindent 6.2.4 \quad Non-Omics data\\
For plant species leaves dataset, S4 and Gap noth correctly select number of cluster $K$ and have $ARI=1$, whereas S4 selects fewer number of features than Gap. For ISOLET dataset, only S4 correctly estimate the number of clusters $K$ with the highest ARI compared to Gap and PS. This demonstrates the capability of S4 not only in genomic field but also applicable to other high-dimensional data to estimate $K$ and $\lambda$ simultaneously.

% \usepackage{multirow}
\begin{table}[]
	\caption{The result for real application}\label{table:Realdataresult} 
	\resizebox{\textwidth}{!}{
		\begin{tabular}{|l|l|l|l|l|l|}
			\hline
			\multirow{2}*{Dataset}        & 	\multirow{2}*{Method}             & 	\multirow{2}*{True K}             & 	\multirow{2}*{K selected}  & \multirow{2}*{\tabincell{c}{Number of \\features selected}} & \multirow{2}*{ARI}  \\ 
			& & & & &\\
			\hline
			\multirow{3}{*}{\cite{verhaak2009prediction}}     & S4                  & \multirow{3}{*}{3} & 3          & 24                          & 0.94 \\
			& Gap Statistic       &                    & 3          & 625                        & 0.89 \\
			& Prediction Strength &                    & 2          & 1680                         & 0.52 \\ \hline
			\multirow{3}{*}{\cite{balgobind2010evaluation}}   & S4                  & \multirow{3}{*}{3} & 3          & 64                          & 0.96    \\ 
			& Gap Statistic       &                    & 3          & 491                         & 0.96 \\ 
			& Prediction Strength &                    & 3          & 61                          & 1    \\ \hline
			\multirow{3}{*}{\cite{kohlmann2008international}} & S4                  & \multirow{3}{*}{3} & 3          & 81                          & 0.97 \\
			& Gap Statistic       &                    & 3          & 330                        & 0.94 \\
			& Prediction Strength &                    & 2          & 1509                        & 0.53 \\ \hline
			\multirow{3}{*}{Mammalian tissue datasets}                            & S4                  & \multirow{3}{*}{4} & 3          & 18                         & 0.65 \\
			& Gap Statistic       &                    & 2          & 54                          & 0.33 \\
			& Prediction Strength &                    & 2          & 1631                        & 0.33 \\ \hline
			\multirow{3}{*}{Rat brain dataset}                                    & S4                  & \multirow{3}{*}{3} & 3          & 79                          & 1    \\
			& Gap Statistic       &                    & 3          & 295                         & 1    \\
			& Prediction Strength &                    & 3          & 1584                       & 1    \\ \hline
			\multirow{3}{*}{Pan-cancer dataset}                                   & S4                  & \multirow{3}{*}{5} & 6          & 46                          & 0.78 \\
			& Gap Statistic       &                    & 2          & 1485                        & 0.22 \\
			& Prediction Strength &                    & 2          & 1297                        & 0.22 \\ \hline
			\multirow{3}{*}{SNP dataset}                                          & S4                  & \multirow{3}{*}{3} & 3          & 5595                        & 0.92 \\
			& Gap Statistic       &                    & 3          & 7397                        & 0.92 \\
			& Prediction Strength &                    & 3          & 10834                       & 0.92 \\ \hline
			\multirow{3}{*}{Plant species leaves dataset}                      & S4                  & \multirow{3}{*}{4} & 4          & 130                        & 1    \\
			& Gap Statistic       &                    & 2          & 71                          & 0.32\\
			& Prediction Strength &                    & 4          & 162                         & 1\\ \hline
			\multirow{3}{*}{ISOLET datasets}                                   & S4                  & \multirow{3}{*}{5} & 5          & 407                          & 0.95  \\
			& Gap Statistic       &                    & 3          & 156                         & 0.53 \\
			& Prediction Strength &                    & 2          & 56                          & 0.53 \\ \hline
	\end{tabular}}
\end{table}

\section{Discussion}
\label{sec:Discussion}
In this paper, we propose S4 method for estimating $K$ in $K$-means and for $K$ and $\lambda$ simultaneously in sparse $K$-means. This issue is particularly important when clustering high-dimensional data. However, to the best of our knowledge, there is no method and evaluation designed for this purpose yet. Our approach utilizes resampling technique to evaluate clustering stability between repeated subsampled data and original data. We hence borrow the notion of sensitivity and specificity to form our target function. We first evaluate S4 method by estimating $K$ for $K$-means without feature selection. We generate 10 different low-dimensional simulation settings and compare S4 with 9 traditional methods including gap statistic and prediction strength. Our method is almost always among the top performers across all settings and shows robust performance.

When estimating both $K$ and $\lambda$ simultaneously, since there is no existing method, we first propose modification of Gap statistic and prediction strength to achieve the goal and then compare the performance with S4. We consider the case when features are independent and the case when features (genes) are correlated. Given our comprehensive simulation studies, we confirm the superior performance of S4 over the other two methods in both clustering accuracy and feature selection. When genes are independent, S4 outperforms Gap and PS especially when number of informative features is relatively small (i.e. Gap and PS tend to over-estimate the number of discriminant features). In the case of correlated gene structure, when the correlation is weak, the performance of Gap and PS drop dramatically, whereas S4 still performs well. In real applications, we evaluated nine datasets and S4 almost always detects fewer number of features with better clustering accuracy. As clustering applications in high-dimensional data becomes more and more prevalent in the future and new methods to simultaneously estimate $K$ and $\lambda$ will be developed, the extensive simulation settings and real applications in this paper can serve as a standard benchmark for evaluation purpose.

%we extended Gap and prediction strength (PS) and compared with S4. In simulation studies, we found that S4 was the only method to achieve near-perfect performance in clustering accuracy and feature selection. Feature selection using Gap and PS generally performed poorly. The result was further confirmed in 9 real application studies including genomic studies and non-genomic studies where s4 either solely selects correct number of cluster K or using smaller number of features to achieve higher or equal ARI. The good performance in non-genomic studies shows the potential for s4 to outperforms beyond genomic field.

Two potential limitations exist for our study. Firstly, since Gap, PS and S4 are all based on resampling approach, they are computationally intensive. In the transcriptomic applications such as the three leukemia studies, it took about 1 hour using 20 computing core of CPU and 128GB RAM to run S4 method if B is 500 and $K$ is selected from 2 to 7, with 30 $\lambda$ for each $K$ , which is affordable even for regular omics studies. Secondly, the performance we have shown is only based on $K$-means and Sparse $K$-means. However, conceptually it can be applied to parameter estimation of other clustering methods with sparse feature selection, such as sparse Gaussian mixture model \citep{pan2007penalized,zhou2009penalized} or sparse Poisson mixture model \citep{witten2011classification}. 
%An R package ``KLCLust'' is publicly available at \url{https://github.com/Statdaydreamer/KLClust_S4}, along with data and code to reproduce all tables and figures in this paper. 
%and meta-analytic sparse $K$-means \citep{huo2016meta}.

\section*{Acknowledgements}
YL, CL and GCT are supported by NIH R01CA190766 and R21LM012752.

%when extended simultaneous estimation of K  and $\lambda$, all the resampling based methodsin Table \ref{Methodintro} are computationally intensive. In the transcriptomic application such as three leukemia stuides, it took about 2 days for a standard PC using 1 computing cores and 16GB RAM to implement S4 method if B is 100 and K is selected from 2 to 7, which is affordable for regular genomic studies. Secondly, we only apply S4 for the $K$-means and sparse $K$-means frameworks. But it conceptually can be used for parameter estimation for any clustering methods such as hierarchical clustering, sparse Gaussian mixture model \citep{pan2007penalized,zhou2009penalized} and meta-analytic sparse $K$-means \citep{huo2016meta}.

\section*{SUPPLEMENTARY MATERIALS}

The illustration of elbow point of within cluster sum of squares(WCSS) is in Figure S1. The illustration of selecting largest $K$ if several $K$ have the same $S_{\rho}(K)$ is in Figure S2. The clarification of trimmed mean using order concordance score is in Figure S3, Figure S4 and Figure S5. The clarification for not estimating both $K$ and $\lambda$ by optimizing the sum of cluster concordance scores and feature concordance scores is in Supplement Simulation S1.

\newpage

%% HERE WE DECLARE THE BIBLIOGRAPHYSTYLE TO USE AND THE BIBLIOGRAPHY DATABASE
\bibliographystyle{apalike}
\bibliography{ref}

\end{document}

%\section{Figure S1: illustration of elbow point for WCSS(K)}

\begin{figure}[h]
	
	\centering
	\subfigure[Three-clusters data]{                    
		\begin{minipage}[A]{0.48\textwidth}
			\centering                                                     
			\includegraphics[width=8cm,height=6cm]{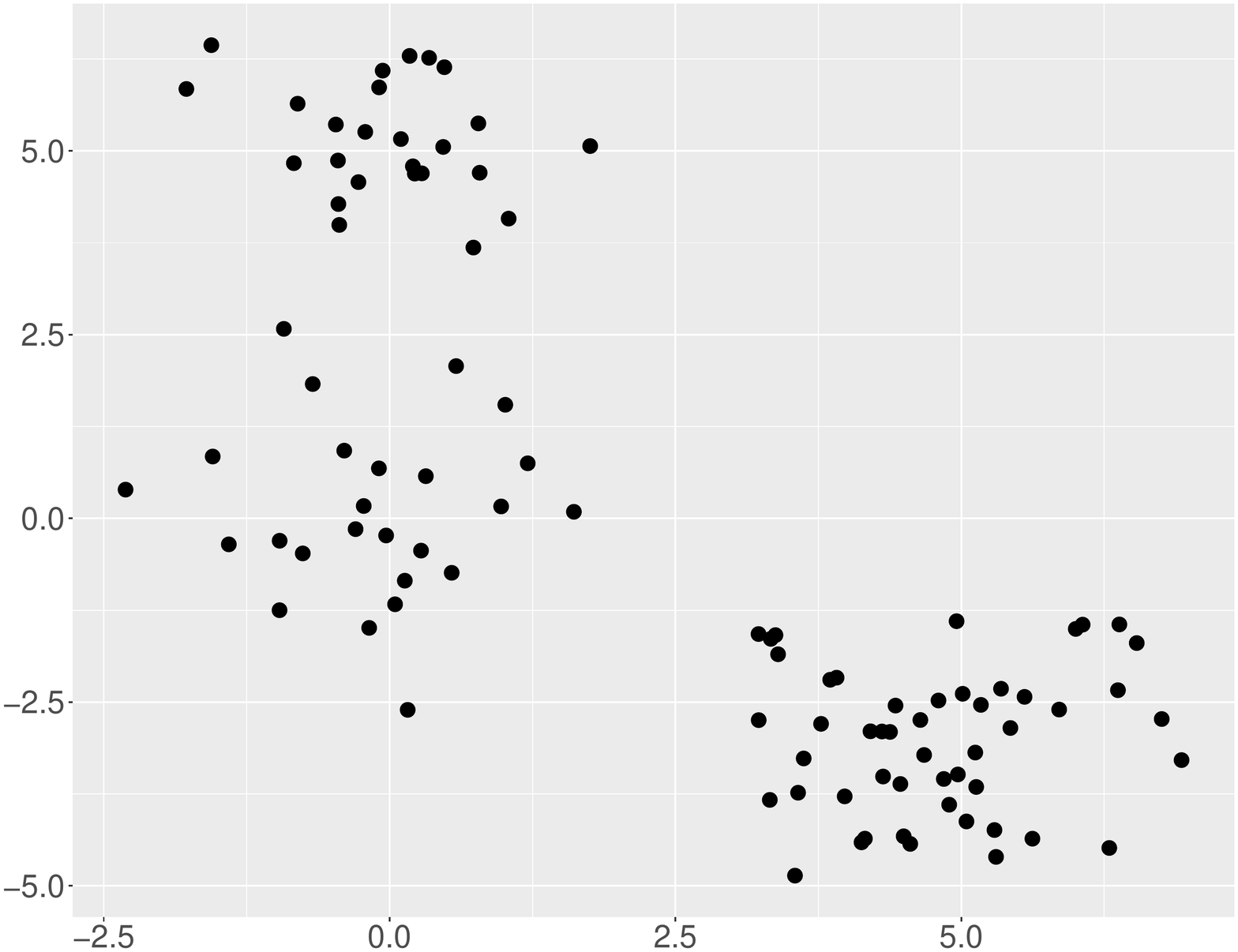}               
	\end{minipage}}
	\subfigure[WCSS of number of clusters K for three-clusters data]{
		\begin{minipage}[B]{0.48\textwidth}
			\centering                                                     
			\includegraphics[width=8cm,height=6cm]{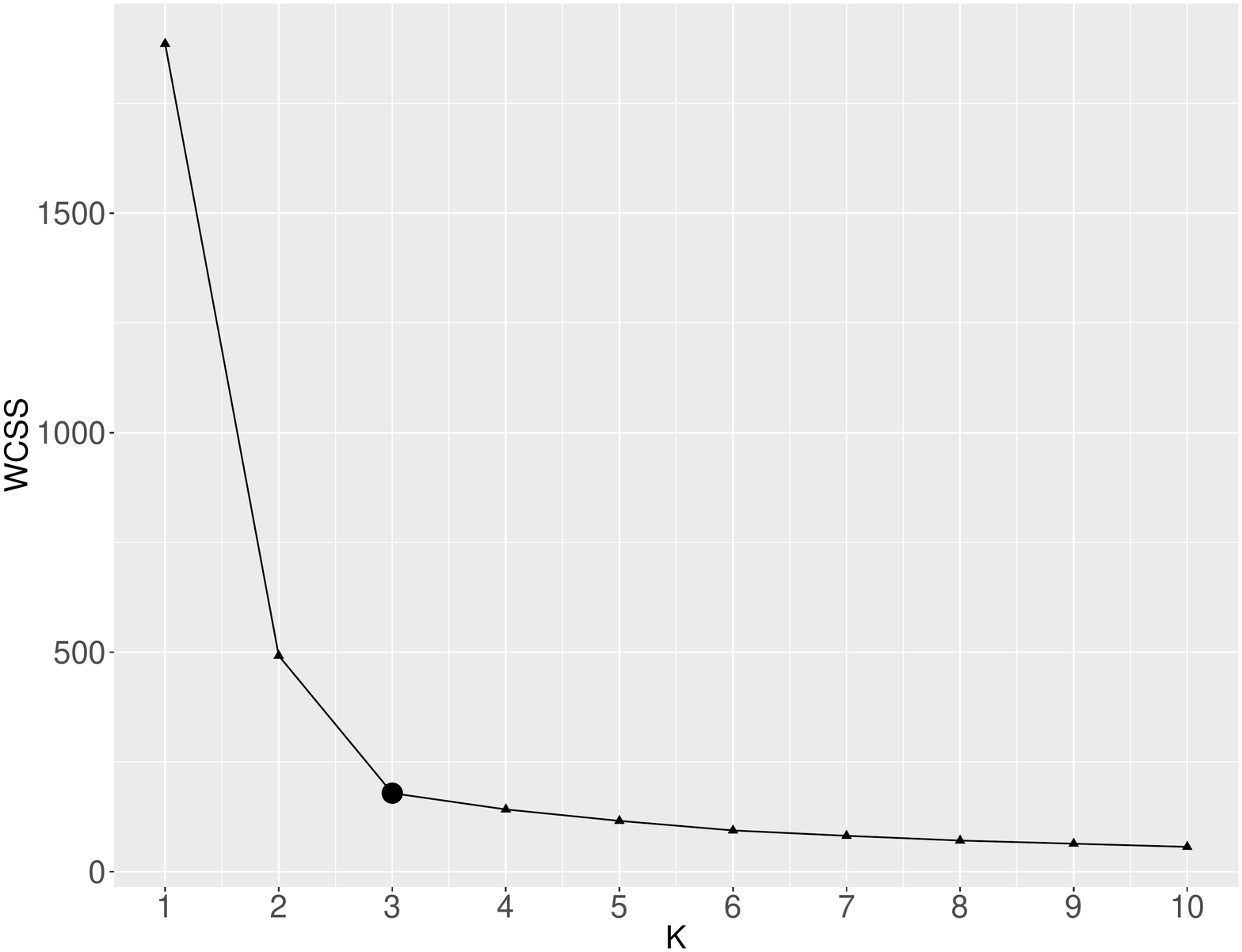}               
		\end{minipage}
	}
	\vfill
	\centering
	\subfigure[Two-clusters data]{                    
		\begin{minipage}{0.48\textwidth}
			\centering                                                     
			\includegraphics[width=8cm,height=6cm]{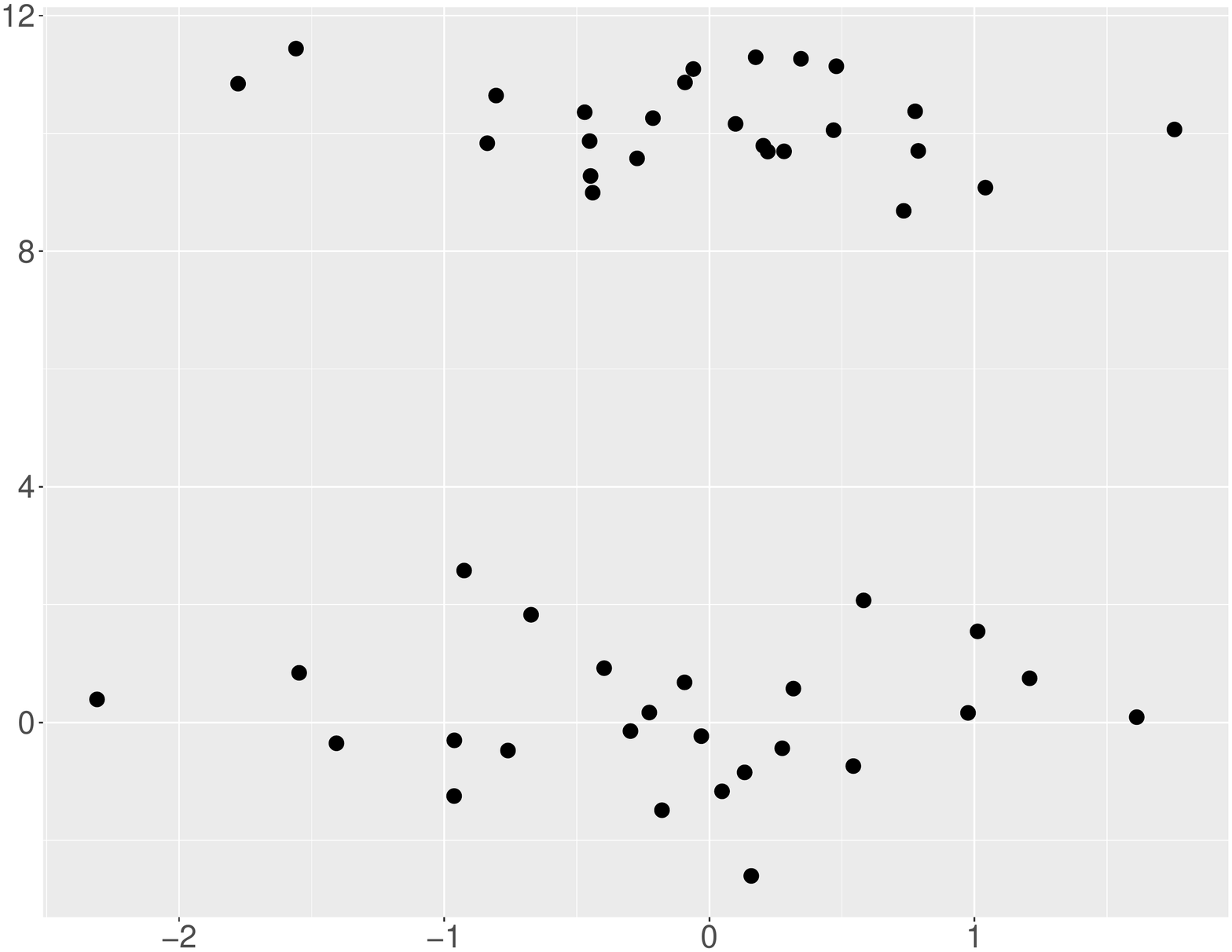}               
	\end{minipage}}
	\subfigure[WCSS of number of clusters K for two-clusters data]{
		\begin{minipage}{0.48\textwidth}
			\centering                                                     
			\includegraphics[width=8cm,height=6cm]{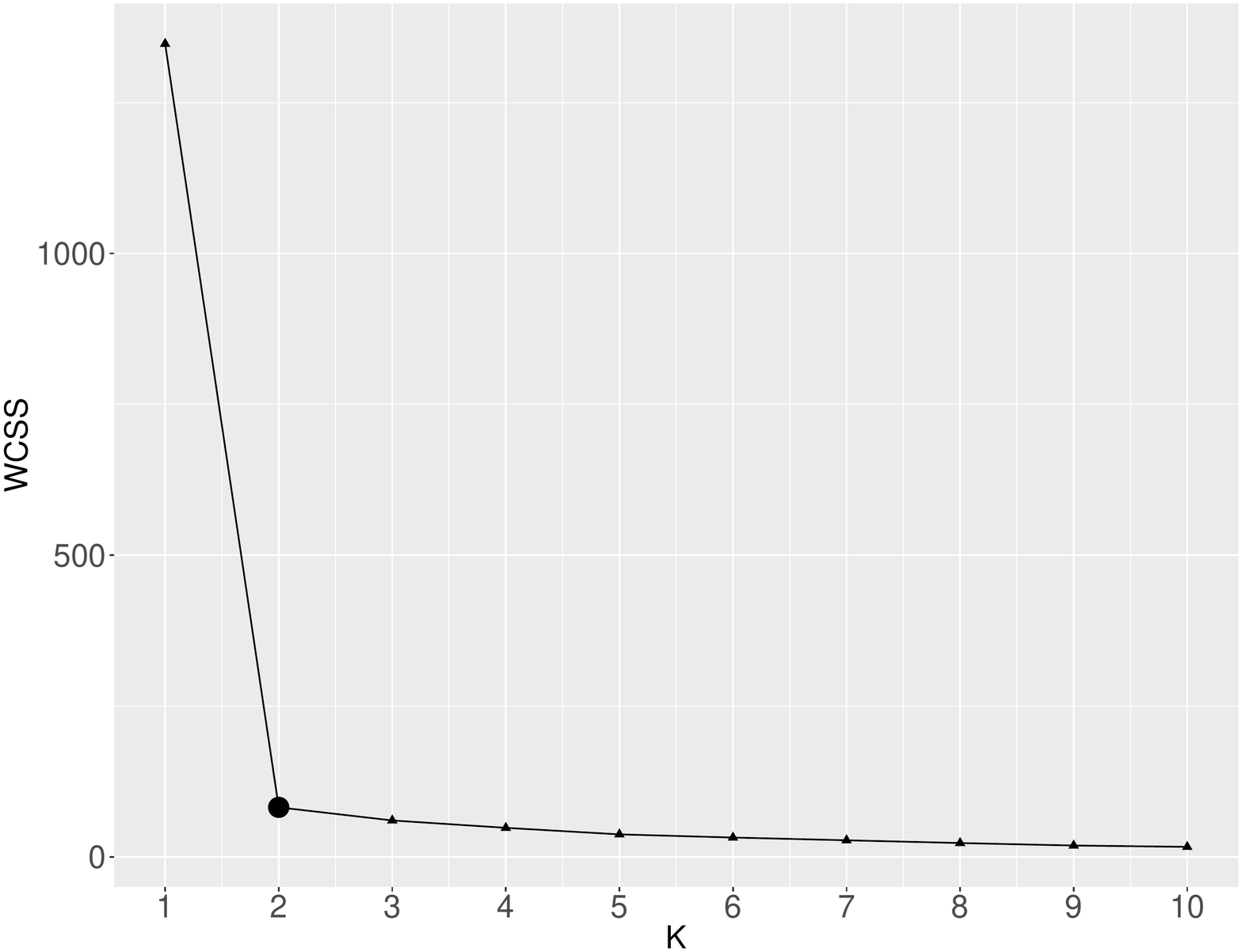}               
		\end{minipage}
	}
	\caption{Illustration of elbow point for WCSS(K)}

\end{figure}

\begin{figure}
	\centering
\includegraphics[width=14cm,height=14cm]{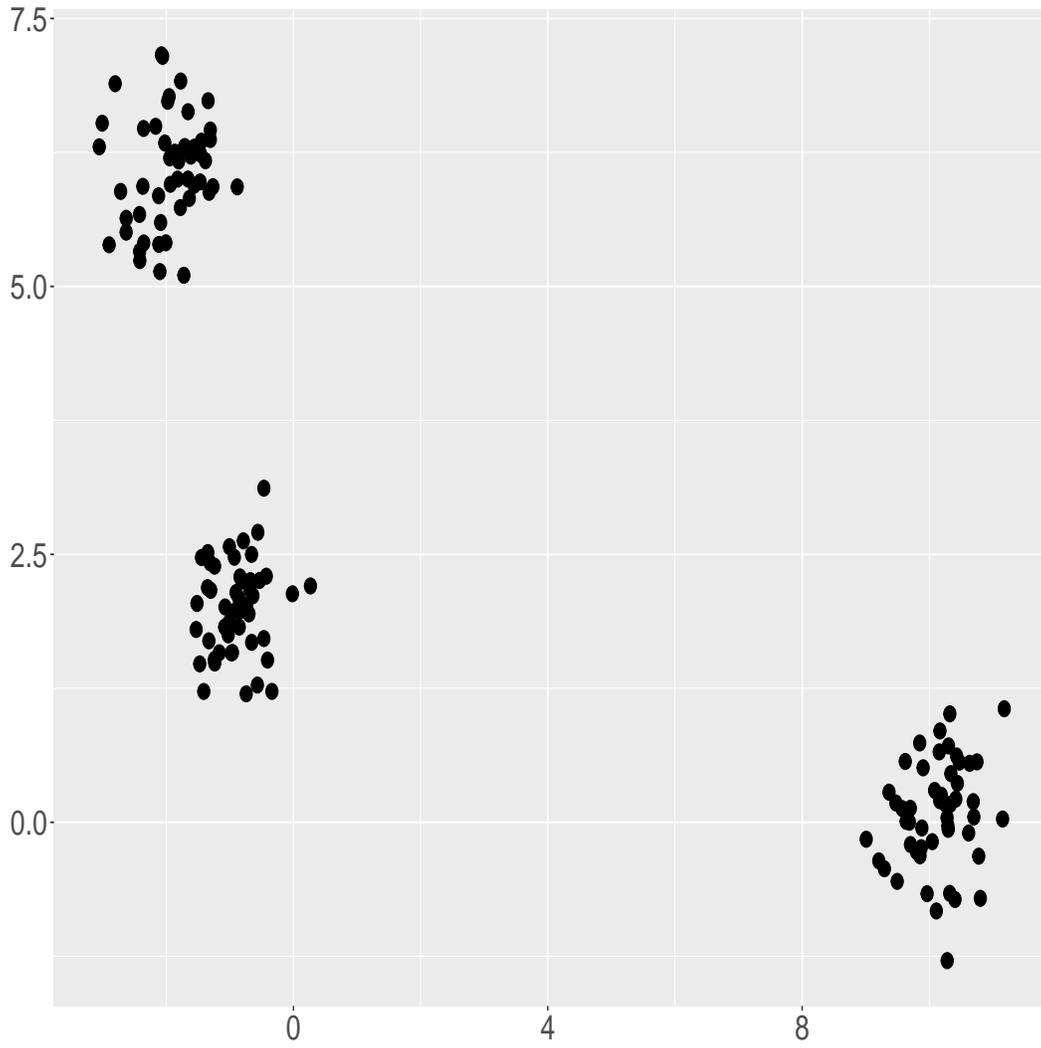} 
\caption{Three clusters where left two are close}
\end{figure}

\begin{figure}
	\centering
	\includegraphics[width=14cm,height=14cm]{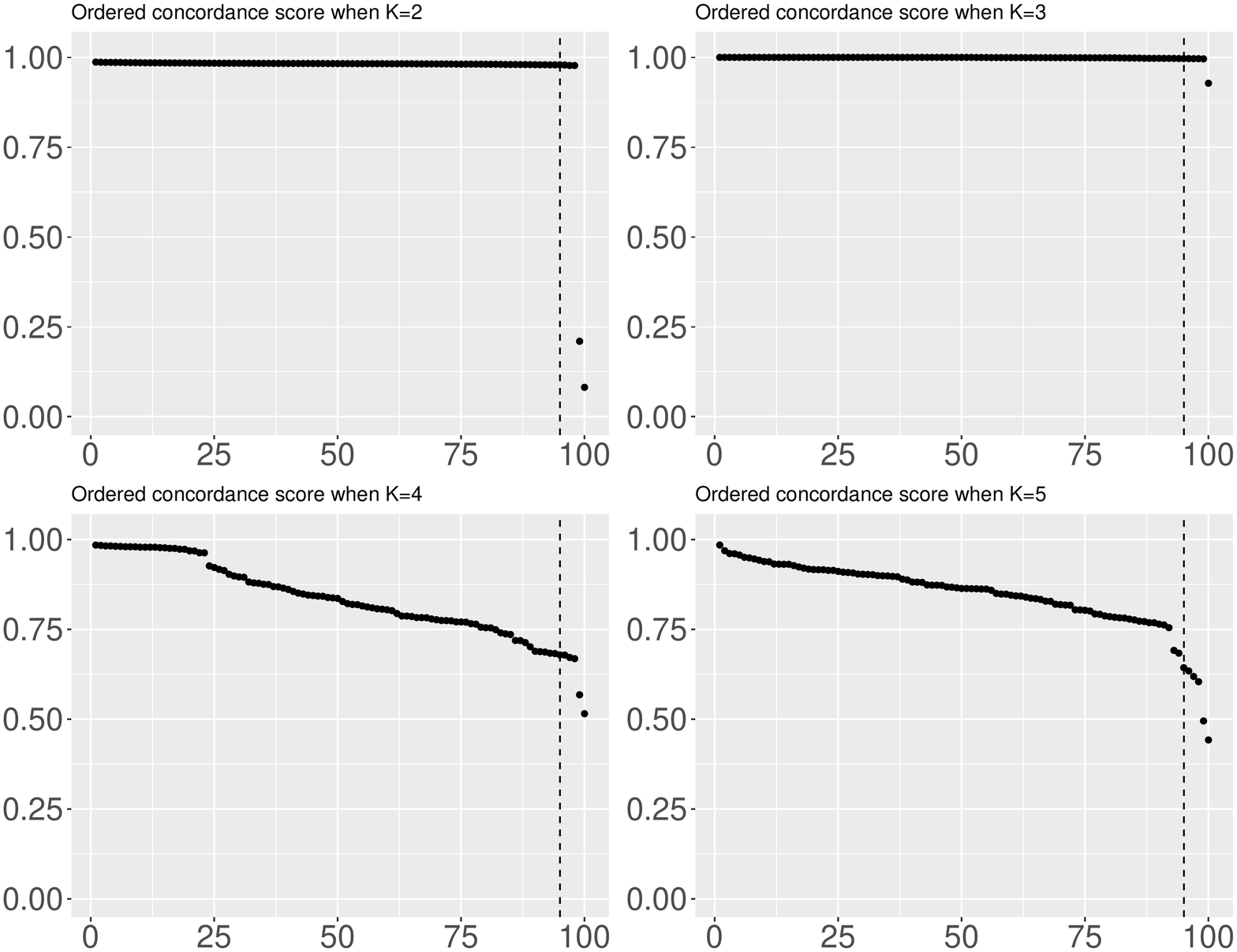}     
	\caption{The ordered concordance score for simulation setting 2  from K=2 to K=5, the dashed line is the 5\% timmed line.}
\end{figure}
\begin{figure}
	\centering
	\includegraphics[width=14cm,height=14cm]{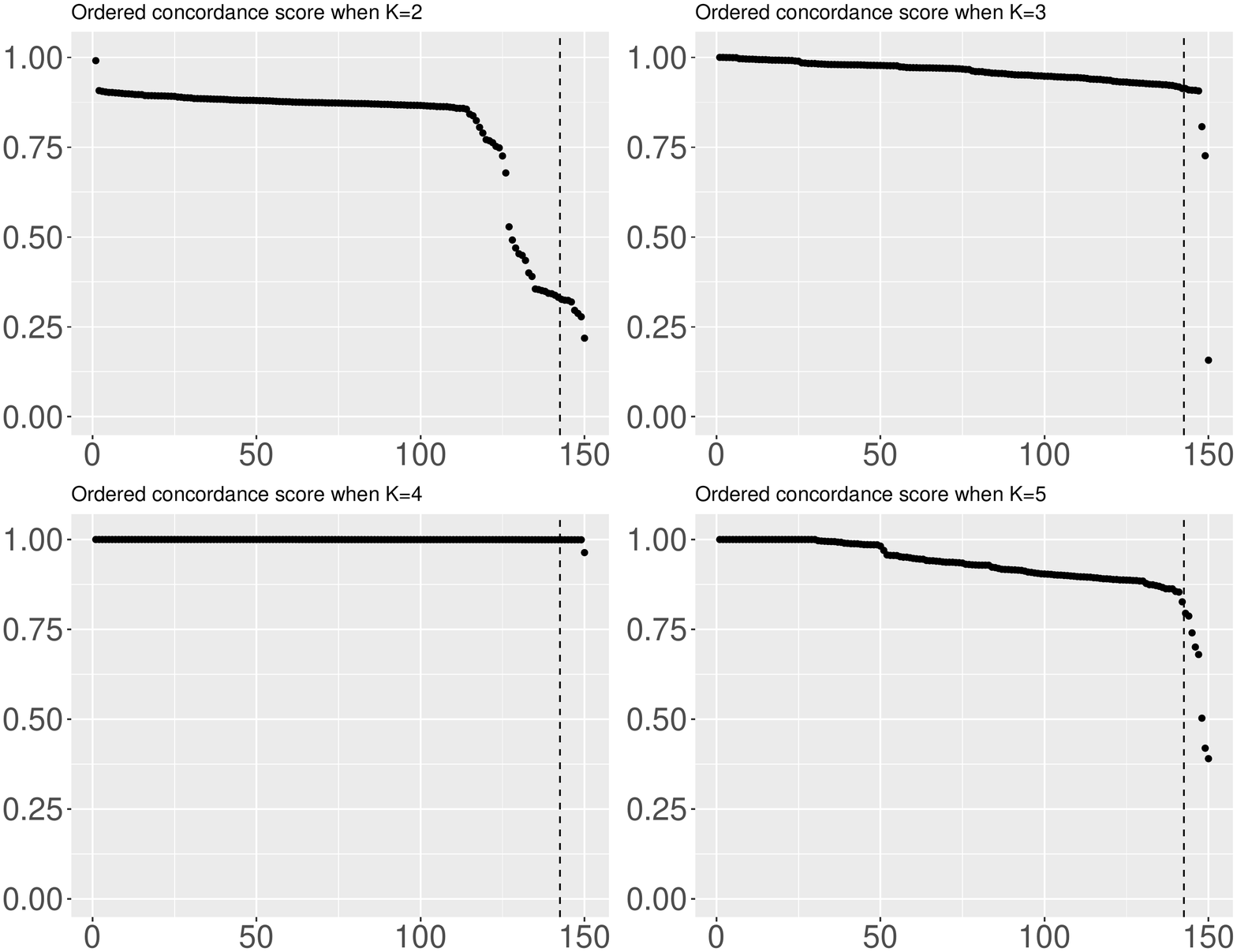}     
	\caption{The ordered concordance score for simulation setting 3  from K=2 to K=5, the dashed line is the 5\% timmed line.}
\end{figure}
\begin{figure}
	\centering
	\includegraphics[width=14cm,height=14cm]{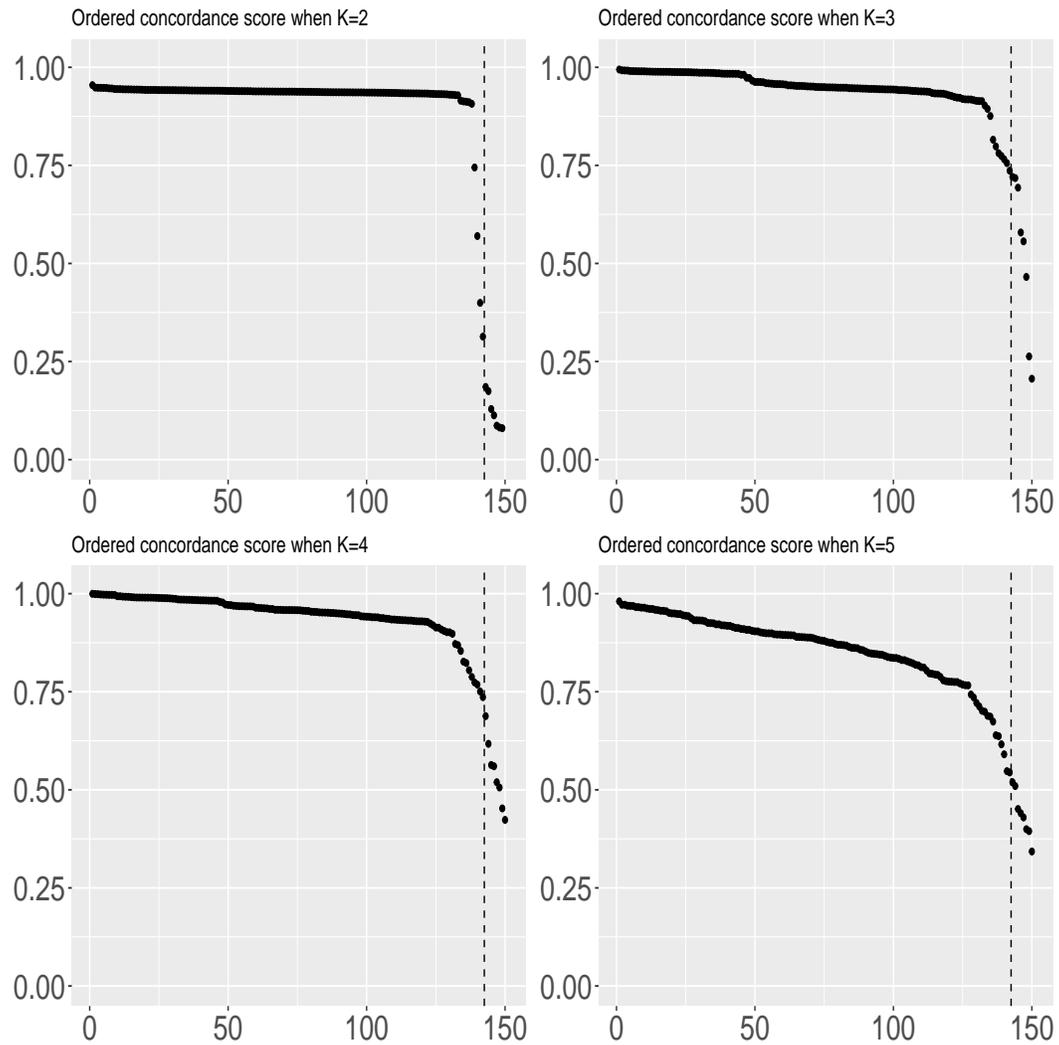}  
	\caption{The ordered concordance score for simulation setting 4  from K=2 to K=5, the dashed line is the 5\% timmed line.}   
\end{figure}

\newpage

\section{Simulation S1: The reason for not estimating $K$ and $\lambda$ by optimizing the sum of cluster concordance score and feature concordance score}	
\noindent \textit{\underline{ }} 

\noindent We here illustrate by simulating data of three clusters where two clusters are obviously closer than another cluster. Denote the whole data matrix by $ X_{n\times p}$ where $n=99$ and $p=300$ and $x_{i,1:j}$ is a vector indicating subject $i$ in from features 1 to feature j.
\begin{itemize}
	\item[1]: Cluster 1: for $1 \leq i \leq 33,$ $x_{i,1:50} \sim mvrnorm(3, I_{50})$, $x_{i,51:150} \sim mvrnorm(0.6, I_{100})$, $x_{i,151:300} \sim mvrnorm(0, I_{150})$, where mvrnorm is abbreviation of multivariate normal distribution.
	\item[2]: Cluster 2: for $1 \leq i \leq 33,$ $x_{i,1:50} \sim mvrnorm(-1, I_{50})$, $x_{i,51:150} \sim mvrnorm(0, I_{100})$, $x_{i,151:300} \sim mvrnorm(0, I_{150})$.
	\item[3]: Cluster3: for $1 \leq i \leq 33,$ $x_{i,1:50} \sim mvrnorm(-1, I_{50})$, $x_{i,51:150} \sim mvrnorm(-1.5, I_{100})$, $x_{i,151:300} \sim mvrnorm(0, I_{150})$.
\end{itemize}	
	
We determine the number of cluster K from 2:7 by S4 method we propose in main section of paper denoted as $S4^*$ and another stategy, denoted as $S4^{**}$, which estimates $K$ and $\lambda$ by optimizing the sum of cluster concordance score and feature concordance score. We do simulation for 50 times and each time we use $B=100$ subsample. The results show that $S4^*$ always choose $K=3$ whereas $S4^{**}$ always choose $k=2$.

The reason for failure of $S4^{**}$ is that in this simulation setting, $S(K=2,\lambda)$ and  $S(K=3,\lambda)$  will both be 1 if $\lambda$ is large enough since the first 150 features well separate three clusters. However, $F(K=3, \lambda)$ is lower than $F(K=2, \lambda)$ since t features $1\sim50$ and features $51\sim150$ both contributes to the clustering if $K=3$, feature selection is unstable compared with $K=2$ where only the features $1\sim50$ strongly contribute to the clustering. Therefore, since $S4^{**}= \arg \max\limits_{K, \lambda} S(K,\lambda)+F(K,\lambda)$, it will choose $K^{**}=2$ where as $S4^*$ choose $K^*=3$ since it only use $S(K,\lambda)$ when estimating $K$.